\documentstyle[epsf]{mn}

\newif\ifAMStwofonts

\def\etal{{et al. \rm}}
\def\ha{H$\alpha$}
\def\has{H$\alpha$ }
\def\he{He\,I $\lambda$6678}
\def\hes{He\,I $\lambda$6678 }

\ifoldfss
  \ifCUPmtlplainloaded \else
    \NewTextAlphabet{textbfit} {cmbxti10} {}
    \NewTextAlphabet{textbfss} {cmssbx10} {}
    \NewMathAlphabet{mathbfit} {cmbxti10} {} 
    \NewMathAlphabet{mathbfss} {cmssbx10} {} 
  \fi
  \ifAMStwofonts
    \ifCUPmtlplainloaded \else
      \NewSymbolFont{upmath} {eurm10}
      \NewSymbolFont{AMSa} {msam10}
      \NewMathSymbol{\upi}     {0}{upmath}{19}
      \NewMathSymbol{\umu}     {0}{upmath}{16}
      \NewMathSymbol{\upartial}{0}{upmath}{40}
      \NewMathSymbol{\leqslant}{3}{AMSa}{36}
      \NewMathSymbol{\geqslant}{3}{AMSa}{3E}

    \fi
  \fi
\fi 

\ifnfssone
  \newmathalphabet{\mathit}
  \addtoversion{normal}{\mathit}{cmr}{m}{it}
  \addtoversion{bold}{\mathit}{cmr}{bx}{it}
  \newmathalphabet{\mathbfit} 
  \addtoversion{normal}{\mathbfit}{cmr}{bx}{it}
  \addtoversion{bold}{\mathbfit}{cmr}{bx}{it}
  \newmathalphabet{\mathbfss} 
  \addtoversion{normal}{\mathbfss}{cmss}{bx}{n}
  \addtoversion{bold}{\mathbfss}{cmss}{bx}{n}
  \ifAMStwofonts
    \ifCUPmtlplainloaded \else
      \UseAMStwoboldmath
      \makeatletter
      \new@mathgroup\upmath@group
      \define@mathgroup\mv@normal\upmath@group{eur}{m}{n}
      \define@mathgroup\mv@bold\upmath@group{eur}{b}{n}
      \edef\UPM{\hexnumber\upmath@group}
      \new@mathgroup\amsa@group
      \define@mathgroup\mv@normal\amsa@group{msa}{m}{n}
      \define@mathgroup\mv@bold\amsa@group{msa}{m}{n}
      \edef\AMSa{\hexnumber\amsa@group}
      \makeatother
      \mathchardef\upi="0\UPM19
      \mathchardef\umu="0\UPM16
      \mathchardef\upartial="0\UPM40
      \mathchardef\leqslant="3\AMSa36
      \mathchardef\geqslant="3\AMSa3E
    \fi
  \fi
\fi 

\ifnfsstwo
  \DeclareMathAlphabet{\mathbfit}{OT1}{cmr}{bx}{it}
  \SetMathAlphabet\mathbfit{bold}{OT1}{cmr}{bx}{it}
  \DeclareMathAlphabet{\mathbfss}{OT1}{cmss}{bx}{n}
  \SetMathAlphabet\mathbfss{bold}{OT1}{cmss}{bx}{n}
  \ifAMStwofonts
    \ifCUPmtlplainloaded \else
      \DeclareSymbolFont{UPM}{U}{eur}{m}{n}
      \SetSymbolFont{UPM}{bold}{U}{eur}{b}{n}
      \DeclareSymbolFont{AMSa}{U}{msa}{m}{n}
      \DeclareMathSymbol{\upi}{0}{UPM}{"19}
      \DeclareMathSymbol{\umu}{0}{UPM}{"16}
      \DeclareMathSymbol{\upartial}{0}{UPM}{"40}
      \DeclareMathSymbol{\leqslant}{3}{AMSa}{"36}
      \DeclareMathSymbol{\geqslant}{3}{AMSa}{"3E}
    \fi
  \fi
\fi 

\ifCUPmtlplainloaded \else
  \ifAMStwofonts \else 
    \def\upi{\pi}
    \def\umu{\mu}
    \def\upartial{\partial}
  \fi
\fi

\title[Large-scale wind structures in OB supergiants]{Large-scale wind structures in OB supergiants: a search for rotationally modulated \has variability\thanks{Based on observations collected at the Vainu Bappu Observatory (Kavalur, India).}}
\author[T. Morel, \etal]
{\parbox{179mm}{\begin{flushleft}
\vspace{-0.5cm}
{\LARGE T. Morel,$^{1,2}$} 
\thanks{e-mail: morel@astropa.unipa.it}
{\LARGE S. V. Marchenko,$^{3}$}
{\LARGE A. K. Pati,$^{4}$}
{\LARGE K. Kuppuswamy,$^{5}$}
{\LARGE M. T. Carini,$^{3}$}
{\LARGE E. Wood$^{3}$}
{\LARGE and R. Zimmerman$^{3}$}
\end{flushleft}
}\vspace*{0.200cm}\\ 
\parbox{159mm}{
$^1$ Istituto Nazionale di Astrofisica, Osservatorio Astronomico di Palermo G. S. Vaina, Piazza del Parlamento 1, I-90134 Palermo, Italy\\
$^2$ Inter-University Centre for Astronomy and Astrophysics (IUCAA), Post Bag 4, Ganeshkhind, Pune, 411 007, India\\
$^3$ Department of Physics and Astronomy, Western Kentucky University, 1 Big Red Way, Bowling Green, KY 42101--3576, USA\\
$^4$ Indian Institute of Astrophysics (IIAP), III block, Koramangala, Bangalore, India\\
$^5$ Vainu Bappu Observatory, Indian Institute of Astrophysics (IIAP), Kavalur, Alangayam 635 701 India}}

\date{Accepted ???.
      Received ???;
      in original form ??? }

\pagerange{\pageref{firstpage}--\pageref{lastpage}}
\pubyear{2003}

\begin{document}

\maketitle

\label{firstpage}

\begin{abstract}
We present the results of a long-term monitoring campaign of the \has line in a sample of bright
OB-supergiants (O7.5--B9) that aims at detecting rotationally modulated changes potentially related to
the existence of large-scale wind structures. A total of 22 objects were monitored during 36 nights spread
over 6 months in 2001--2002. Coordinated broad-band photometric observations were also obtained for
some targets. Conspicuous evidence for variability in \has is found for the stars displaying a feature contaminated by wind emission.  Most changes take place on a daily time-scale, although hourly variations are also occasionally detected. Convincing evidence for a cyclical pattern of variability in \has has been found in 
2 stars: HD 14134 and HD 42087 (periodic signals are also detected in other stars, but independent confirmation
is required). Rotational modulation is suggested from the similarity between the observed recurrence time-scales (in the range
13--25 days) and estimated periods of stellar rotation. We call attention to the atypical case of HD 14134 which exhibits a clear 12.8-d periodicity both in
the photometric and in the spectroscopic data sets. This places this object among a
handful of early-type stars where one may observe a clear link
between extended wind structures and photospheric disturbances. Further
modelling may test the hypothesis that azimuthally-extended wind streams are responsible for the patterns
of spectral variability in our target stars.
\end{abstract}

\begin{keywords}
stars: early-type -- stars: supergiants -- stars: emission-line -- stars: winds, outflows -- stars: rotation.
\end{keywords}

\section{Introduction}
One of the major legacies of the {\it International Ultraviolet Explorer (IUE)} was to establish
the ubiquity of UV line-profile variability in OB stars.  The pattern of variability predominantly
consists of extra absorption features (the so-called 'discrete absorption components';
hereafter DACs) moving blueward across the (unsaturated) absorption troughs of P Cygni lines
(e.g., Kaper \etal 1996). The repetitive nature of DACs and the fact that the recurrence time-scale is
commensurate with the (otherwise ill-defined) stellar rotational period suggest that rotation plays a key role in inducing the patterns of variability (e.g., Kaper \etal 1999).
UV time-series with a long temporal baseline (i.e., sampling several rotational periods) have
revealed that OB stars exhibit a great diversity of behaviours that cannot always be strictly related to the
stellar rotation period (e.g., Howarth, Prinja \& Massa 1995; Prinja, Massa \& Fullerton 1995; Prinja,
Massa \& Fullerton 2002). Nevertheless, this global phenomenon has been successfully interpreted in several cases
in the framework of a model in which
the changes are caused by the rotational modulation of large-scale, azimuthally-extended wind streams
(e.g., Fullerton \etal 1997).

The hydrodynamical simulations of Cranmer \& Owocki (1996) suggest that large-scale photospheric
perturbations (e.g., bright 'spots' causing a local enhancement of the radiative driving force) may lead to the development of these structures. Although the underlying physical mechanism remains elusive,
pulsational instability can be invoked as a trigger of the wind variability (e.g., Fullerton, Gies \& Bolton 1996).
Large-scale or 'surface' magnetic field structures could be a viable alternative in view of the growing evidence for  (weak) magnetic fields in OB stars (e.g., Henrichs, Neiner \& Geers 2003).
Photospheric perturbations arising from these two phenomena are likely to have a significant impact on
the global wind morphology (e.g., ud-Doula \& Owocki 2002).

With the demise of {\it IUE}, further progress on this issue is likely
to stem in the foreseeable future from time-resolved studies of line variability in the optical. Firstly, intensive UV observations have been carried out over several rotation
periods for few stars (Massa \etal 1995; Prinja \etal 1998; de Jong \etal 2001). For practical reasons, an optical survey is much more
amenable at collecting high-quality observations covering several cycles for a large, representative sample
of OB stars. Secondly, in contrast
to the UV resonance lines, optical recombination lines are (at least partly) formed in OB supergiants in the dense, strongly accelerating
part of the outflow close to the stellar core. In virtue of their density-squared dependence, they are good probes of the large density gradients expected to prevail in the co-rotating interaction regions
 (Cranmer \& Owocki 1996). Revealing a cyclical pattern of variability in these lines would give further credence to the idea
that the origin of anisotropic outflows in OB stars is deep-seated and causally linked to processes taking place
at the stellar photosphere. Here we discuss the temporal behaviour of \ha, which is typically formed at $r$$\sim$ 1.5 $R_{\star}$ in O-type stars (e.g., Prinja, Fullerton \& Crowther 1996). Although the line-formation region is likely to be more extended in B-type supergiants, this transition still probes the few inner radii of the stellar wind.

Previous surveys have been very useful in documenting \has line-profile variability in OB stars (e.g., Ebbets 1982), but frequently suffered from a poor temporal sampling  hampering the study of the
line-profile variations on a rotational time-scale (up to several weeks in the case of B supergiants). Despite the
persistent and numerous searches for optical line-profile variability in OB stars, evidence for
rotational modulation has been found in only few cases (e.g., Moffat \& Michaud 1981; Kaper \etal 1997, 1998).
In the UV and optical domains the time-scales associated with the changes are often identical,
within the uncertainties (e.g., Howarth \etal 1995; Kaper \etal 1999; de Jong \etal 2001; Kaufer, Prinja \& Stahl 2002).
This suggests that \has variability and DACs diagnose the same underlying physical phenomenon.

The long-term monitoring of the \has line presented in this paper aims at better establishing the incidence of rotationaly modulated variability in OB
supergiants. To potentially link the photospheric and wind activity, coordinated broad-band photometic observations were also carried out for a subset of this sample. Our survey primarily targets early B-type stars and
can thus be regarded as complementary to others that concentrate on O-type (Kaper 1998; Kaper \etal 1998) or
late B- to early A-type supergiants (Kaufer \etal 1996). The results presented in the following supersede previous
preliminary reports (Morel \etal 2004a, b).

\section{Target selection and temporal sampling of the observations}
Our targets are drawn from a magnitude-limited ($V$$<$7.5) sample of early-type supergiants (O7.5--B9)
with evidence for an emission-like \has profile or, in the case of stars with an absorption feature,
for previous claims of a wind-related variability (e.g., filling-in of the profile). Known close binaries were discarded from this list in order to avoid contamination from the wind collision effects (e.g., Thaller \etal 2001). The wide binary orbit of HD 37742 rules out the formation of a bow shock created from the interaction with an O-type companion (Hummel \etal 2000). In our data only HD 13854 and HD 47240 present significant velocity variations of the photospheric He\,I $\lambda$6678.15 line which might be attributed to binary motion (see below). Dwarfs and giant stars were not included in this list, as \has is of photospheric
origin in the vast majority of these objects. The only exception was the O7.5 giant HD 24912 ($\xi$ Per) because
of previous claims of a cyclical, wind-related pattern of variability in \has (Kaper \etal 1997; de Jong \etal 2001). For the other stars in our list, no clear evidence for cyclical line-profile variations in optical wind lines was found prior to our survey. Some physical parameters of our program stars are given in Table \nolinebreak 1.

\begin{table*}
\centering
\caption{Physical parameters of the program stars: spectral types and wind terminal velocities,
$v_{\infty}$, from Howarth \etal (1997) and references therein; effective temperatures from the calibration of Humphreys \&
MacElroy (1984); stellar luminosities calculated from the calibrated $M_V$ values and bolometric
corrections of Humphreys \& MacElroy (1984); stellar masses from Lamers (1981) (the numbers
in brackets are estimates based on his values for similar spectral types and luminosity classes). This source was used because it constitutes the most homogeneous database, but we warn the reader that these values may be 
significantly revised (see, e.g., Herrero, Puls \& Najarro 2002);
stellar radii determined from the effective temperatures and luminosities; mass-loss rates, $\dot{M}$,
calculated from the empirical relation for galactic OBA supergiants of Lamers \& Cassinelli (1996)
(the value for HD 24912 was derived from their relation for giants); projected rotational velocities, $v \sin i$,
from Howarth \etal (1997); critical rotational velocities, $v_{\rm crit}$, calculated following Howarth
\& Prinja (1989) (the values in brackets are poorly determined because of an uncertain stellar mass).}
\hspace*{-1.5cm}
\begin{tabular}{lclcccccccc}
\hline
Name & $V$ & Spectral type & $T_{{\rm eff}}$ & $\log(L_{\star}/L_{\odot})$ & $M_{\star}$ & $R_{\star}$ & $\dot{M}$ & $v_{\infty}$ & $v \sin i$ & $v_{\rm crit}$ \\
     &(mag) &                & (K)           &        & (M$_{\odot}$) & (R$_{\odot}$) & (10$^{-6}$ M$_{\odot}$ yr$^{-1}$) & (km s$^{-1}$) &  (km s$^{-1}$) &  (km s$^{-1}$)\\\hline 
HD 13854                   & 6.49 & B1 Iab            & 20,260 & 5.18 & (22) & 32 & 0.76 &  920 & 97 & (328)\\
HD 14134 (V520 Per)        & 6.55 & B3 Ia             & 16,300 & 5.24 & 24   & 52 & 1.45 &  465 & 66 & 265  \\
HD 14818 (10 Per)          & 6.26 & B2 Ia             & 18,000 & 5.40 & (30) & 52 & 2.13 &  565 & 82 & (293)\\
HD 21291                   & 4.25 & B9 Ia$^{a}$       & 10,250 & 4.86 & 19   & 85 &      &      & 30$^{b}$ & 195  \\
HD 24398 ($\zeta$ Per)     & 2.88 & B1 Ib             & 20,260 & 4.90 & 21   & 23 & 0.23 & 1295 & 67 & 396  \\
HD 24912 ($\xi$ Per)       & 4.04 & O7.5 III (n)((f)) & 35,400 & 5.40 & 28   & 13 & 0.22 & 2330 & 213& 550  \\
HD 30614 ($\alpha$ Cam)    & 4.30 & O9.5 Ia           & 29,900 & 5.86 & 40   & 32 & 5.03 & 1590 & 129& 350  \\
HD 31327                   & 6.09 & B2 Ib             & 18,000 & 4.76 & (20) & 25 &      &      & 60 & (377)\\
HD 37128 ($\epsilon$ Ori)  & 1.70 & B0 Ia             & 28,600 & 5.78 & 38   & 32 & 3.15 & 1910 & 91 & 362  \\
HD 37742 ($\zeta$  Ori A)  & 1.7  & O9.7 Ib           & 29,250 & 5.40 & 25   & 20 & 1.05 & 1860 & 124& 421  \\
HD 38771 ($\kappa$ Ori)    & 2.04 & B0.5 Ia           & 23,100 & 5.50 & 35   & 35 & 1.37 & 1525 & 83 & 379  \\
HD 41117 ($\chi^2$ Ori)    & 4.64 & B2 Ia             & 18,000 & 5.40 & 32   & 52 & 2.36 &  510 & 72 & 305  \\
HD 42087 (3 Gem)           & 5.76 & B2.5 Ib           & 17,150 & 4.72 & (19) & 26 & 0.20 &  735 & 71 & (359)\\
HD 43384 (9 Gem)           & 6.29 & B3 Iab            & 16,300 & 4.88 & (22) & 35 & 0.30 &  760 & 59 & (331)\\
HD 47240                   & 6.18 & B1 Ib             & 20,260 & 4.90 & (21) & 23 & 0.31 &  960 & 103& (396)\\
HD 52382                   & 6.51 & B1 Ib             & 20,260 & 4.90 & (21) & 23 & 0.33 &  900 & 74 & (396)\\
HD 53138 ($o^2$ CMa)       & 3.00 & B3 Ia             & 16,300 & 5.24 & 23   & 52 & 0.78 &  865 & 58 & 258  \\
HD 58350 ($\eta$ CMa)      & 2.40 & B5 Ia             & 13,700 & 5.02 & 20   & 58 & 0.91 &  320 & 50 & 238  \\
HD 91316 ($\rho$ Leo)      & 3.84 & B1 Iab            & 20,260 & 5.18 & 22   & 32 & 0.63 & 1110 & 75 & 328  \\
HD 119608                  & 7.50 & B1 Ib             & 20,260 & 4.90 & (21) & 23 & 0.34 &  880 & 74 & (396)\\
HD 151804                  & 5.24 & O8 Iaf            & 33,500 & 6.06 & (60) & 32 & 11.3 & 1445 & 104& (416)\\
HD 152236 ($\zeta^1$ Sco)  & 4.77 & B0.5 Ia+          & 19,700$^{c}$ & 6.05$^{c}$ & (35) & 91$^{c}$ & 5.36 &  390 & 74 & (99)\\\hline
\end{tabular}\\
\begin{flushleft}
$^{a}$ From the SIMBAD database.\\
$^{b}$ From Abt, Levato \& Grosso (2002).\\
$^{c}$ From Rivinius \etal (1997).
\end{flushleft}
\end{table*}

The basic requirement of this survey was to achieve a time sampling allowing a proper assessment
of the variability pattern on a rotational time-scale (typically 1--2 weeks). Although the rotational periods of our program stars are unknown, the critical and projected rotational velocities can be used, along with the estimates of the stellar radii,
to set a lower and an upper limit on this quantity. As can be seen in Table 2, the estimated rotational periods
span a wide range of values (1.2--144 d). Our 22 program stars were monitored during a total of 36
nights spread over 6 months, with 1--2 spectra obtained every night. This long-term monitoring was
necessary to cover at least one rotational cycle for the slowly rotating objects. We also obtained
intranight observations, expecting  significant hourly changes in some cases (e.g., HD 24912). In general,
this temporal sampling appears to be adequate for our purpose (see Table 3). For 5 stars, however, the observations were too
sparse to reliably investigate a potential cyclical behaviour (HD 14818, HD 21291, HD 119608, HD 151804,
and HD 152236). No period search was performed in this case (see Section 5.2), but a qualitative description of the changes is presented in Section 4.

\begin{table*}
\centering
\caption{Summary of period search (all periods in days). $\cal{P}_{\rm min}$ and $\cal{P}_{\rm max}$:
lower and upper limits on the stellar rotational period
(determined from the critical and projected rotational velocities, along with an estimate of the stellar radius);
$\cal{P}$(phot): periodicities in the {\it Hipparcos} data (var: stars displaying photometric variability but
without evidence for periodicity; non-var: stars without significant variability). The light curves are
shown in Fig.2 (and in Fig.5 for HD 14134).; $\cal{P}$(literature): Literature values for the periodicities in \has (Kaper \etal 1997, 1998; de Jong \etal 2001; Markova 2002).
The numbers in italics refer to the periodicities in the UV wind line profiles (Kaper \etal 1999; de Jong \etal 2001). Multiple values refer to periods derived at different epochs. Numbers in brackets indicate that the periodicity found is longer that the time span
of the {\it IUE} observations.
For the \has periodicities derived from our observations, $\cal{P}$(\ha), '2D'
and 'EW' indicate that the period is present in the \has pixel-to-pixel or EW time-series, respectively.}
\hspace*{-1.2cm}
\begin{tabular}{lccccc}
\hline
Name   & $\cal{P}_{\rm min}$ & $\cal{P}_{\rm max}$ & $\cal{P}$(phot) & $\cal{P}$(literature)$^a$ & $\cal{P}$(\ha)$^b$ \\\hline
HD 13854                  & (4.9) & 16.5 & 5.644       &                                             & 1.047 (EW) \\
HD 14134 (V520 Per)       & 10    & 40.1 & 12.823      &                                             & 22.2 (2D), 12.5 (2D and EW)\\
HD 14818 (10 Per)         & (8.9) & 31.8 & 2.754       &                                             & No period search performed\\
HD 21291                  & 22    & 144  & 26.76       & [1,2,3]                                     & No period search performed \\ 
HD 24398 ($\zeta$ Per)    & 2.9   & 17.3 & non-var     &                                             & No periods found \\ 
HD 24912 ($\xi$ Per)      & 1.2   & 3.17 & non-var     & 2.1$\pm$0.1, 1.96$\pm$0.11, 2.086  [4,5,6]  & 2.197 ? (2D and EW)\\
                          &       &      &             & {\it 1.9$\pm$0.3, 1.0$\pm$0.1, 2.0$\pm$0.2} & \\
HD 30614 ($\alpha$ Cam)   & 4.6   & 12.5 & var         & $\sim$5.6, $\sim$7, $\sim$10 [5,6,7,8,9]    & No periods found  \\
HD 31327                  & (3.3) & 20.8 & non-var     &                                             & No periods found  \\
HD 37128 ($\epsilon$ Ori) & 4.4   & 17.6 & var         & [8,10]                                      & 18.2, 0.781 (both 2D)  \\
HD 37742 ($\zeta$  Ori A) & 2.3   & 7.97 & 1.407       & $\sim$6 [5,6,8]                             & $\ga$80 (2D), 50 (2D), 13.3 (EW) \\
                          &       &      &             & {\it (6.3$\pm$2.1), (6.1$\pm$2.7)}          & 1.136 (2D), 0.877 (2D)\\ 
HD 38771 ($\kappa$ Ori)   & 4.7   & 21.4 & var         & [7,8,11]                                    & 4.76, 1.047 (both 2D)   \\
HD 41117 ($\chi^2$ Ori)   & 8.6   & 36.3 & 2.869       & [1,8,12,13]                                 & $\ga$80 (2D), 40 (2D and EW), 0.957 (2D), 0.922 (2D)\\
HD 42087 (3 Gem)          & (3.7) & 18.5 & 6.807       &                                             & 25 (2D and EW) \\
HD 43384 (9 Gem)          & (5.3) & 29.7 & 13.7        &                                             & No periods found  \\
HD 47240                  & (2.9) & 11.3 & 2.7424      &                                             & No periods found \\
HD 52382                  & (2.9) & 15.7 & var         &                                             & No periods found  \\
HD 53138 ($o^2$ CMa)      & 10    & 45.7 & var         & [8,14]                                      & No periods found  \\
HD 58350 ($\eta$ CMa)     & 12    & 58.2 & var         & [8]                                         & No periods found  \\ 
HD 91316 ($\rho$ Leo)     & 4.9   & 21.3 & var$^c$     & [12,13,15]                                     & No periods found  \\
HD 119608                 & (2.9) & 15.7 & non-var     &                                             & No period search performed  \\
HD 151804                 & (3.9) & 15.5 & non-var$^d$ & 7.3 [6,16]                                  & No period search performed  \\
HD 152236 ($\zeta^1$ Sco) & (46)  & 62.3 & var$^e$     & [17,18]                                     & No period search performed  \\\hline
\end{tabular}\\
\begin{flushleft}
$^a$ The numbers in square brackets refer to previous investigations of \has variability in our program stars: [1] Rosendhal (1973); [2] Denizman \& Hack (1988); [3] Zeinalov \& Rzaev
(1990); [4] de Jong \etal (2001); [5] Kaper \etal (1997); [6] Kaper \etal (1998); [7] Ebbets (1980);
[8] Ebbets (1982); [9] Markova (2002); [10] Cherrington (1937); [11] Rusconi \etal (1980); [12] Underhill (1960); [13] Underhill (1961);
[14] van Helden (1972); [15] Smith \& Ebbets (1981); [16] Prinja \etal (1996);
[17] Sterken \& Wolf (1978); [18] Rivinius \etal (1997).\\
$^b$ The typical uncertainties for the quoted periods have been derived from the FWHM of the peaks in the PS and amount to: 20--30 days ($\cal{P}$$\ga$50 d), 1--4 days (10$\la$$\cal{P}$ (d)$\la$25), 0.01--0.2  days (1$\la$$\cal{P}$ (d)$\la$5), and 0.01  days ($\cal{P}$$\la$1 d).\\
$^c$ Multi-periodic variable according to Koen (2001).\\
$^d$ Photometric variable according to Balona (1992), and references therein.\\
$^e$ Irregular, multi-periodic photometric variable (Sterken, de Groot \& van Genderen 1997).
\end{flushleft}
\end{table*}

\begin{table*}
\centering
\caption{Temporal sampling of the spectroscopic observations. $N$: total number of spectra obtained;
$\Delta T$: total time span of the observations; Min($\Delta t$), Mean($\Delta t$), and Max($\Delta t$):
minimum, mean, and maximum time intervals between consecutive spectra, respectively.}
\begin{tabular}{llccccccc}
\hline
Name & Run$^{a}$ & Nb. nights & $N$ & $<$$S/N$$>$ & $\Delta T$ & Min($\Delta t$) & Mean($\Delta t$) & Max($\Delta t$)\\
                          &         &    &    &  & (d) & (d) & (d) & (d) \\\hline   
HD 13854                  & A,B,C   & 16 & 19 & 210 & 71.1 & 0.07 & 3.95 & 24.0\\
HD 14134 (V520 Per)       & A,B,C   & 17 & 22 & 205 & 67.1 & 0.07 & 3.19 & 24.0\\
HD 14818 (10 Per)         & C       & 7  & 7  & 170 & 12.0 & 0.93 & 2.01 & 4.06\\
HD 21291                  & C       & 7  & 7  & 190 & 8.01 & 0.93 & 1.33 & 2.97\\
HD 24398 ($\zeta$ Per)    & A,B     & 9  & 13 & 305 & 38.8 & 0.07 & 3.24 & 24.9\\
HD 24912 ($\xi$ Per)      & A,B,C   & 17 & 23 & 250 & 75.8 & 0.07 & 3.45 & 26.0\\
HD 30614 ($\alpha$ Cam)   & A,B,C   & 16 & 19 & 240 & 70.9 & 0.07 & 3.94 & 26.0\\
HD 31327                  & A,B     & 10 & 12 & 265 & 38.9 & 0.10 & 3.54 & 23.0\\
HD 37128 ($\epsilon$ Ori) & A,B,C,D & 17 & 21 & 335 & 129  & 0.05 & 6.44 & 48.9\\
HD 37742 ($\zeta$  Ori A) & A,B,C,D & 23 & 27 & 275 & 133  & 0.10 & 5.10 & 48.9\\
HD 38771 ($\kappa$ Ori)   & A,B,C,D & 19 & 22 & 280 & 130  & 0.15 & 6.17 & 49.8\\
HD 41117 ($\chi^2$ Ori)   & A,B,C,D & 22 & 25 & 220 & 131  & 0.11 & 5.44 & 49.9\\
HD 42087 (3 Gem)          & A,B,C,D & 20 & 20 & 210 & 135  & 0.86 & 7.09 & 49.8\\
HD 43384 (9 Gem)          & A,C,D   & 17 & 17 & 220 & 133  & 0.87 & 8.29 & 51.9\\
HD 47240                  & A,C,D   & 16 & 16 & 195 & 133  & 0.91 & 8.85 & 51.9\\
HD 52382                  & A,C,D   & 17 & 17 & 180 & 164  & 0.85 & 10.2 & 56.0\\
HD 53138 ($o^2$ CMa)      & A,B,C,D & 20 & 22 & 230 & 164  & 0.10 & 7.79 & 50.9\\
HD 58350 ($\eta$ CMa)     & C,D     & 14 & 17 & 210 & 104  & 0.08 & 6.48 & 48.8\\
HD 91316 ($\rho$ Leo)     & A,B,C,D & 19 & 20 & 200 & 162  & 0.07 & 8.50 & 48.8\\
HD 119608                 & C,D     & 7  & 7  & 135 & 104  & 1.09 & 17.3 & 55.9\\
HD 151804                 & D       & 7  & 7  & 125 & 42.0 & 0.99 & 6.99 & 30.9\\
HD 152236 ($\zeta^1$ Sco) & D       & 6  & 6  & 165 & 42.0 & 1.02 & 8.39 & 31.0\\\hline
\end{tabular}\\
\begin{flushleft}
$^{a}$ A: 2001 Nov. 18--Dec. 04; B: 2001 Dec. 27--28; C: 2002 Jan. 21--Feb. 03; D: 2002 Mar. 24--May 07.
\end{flushleft}
\end{table*}

\section{Observations and reduction procedure}
\subsection{Spectroscopy}
Our observations were obtained during 4 observing runs conducted between 2001 November and 2002
May at the 40-inch telescope of the Vainu Bappu Observatory (Kavalur, India). The spectograph was equipped
with a TK 1k CCD chip. We used the 651 l/mm grating blazed in the 1st order at 7500 \AA, which yields at \has a
reciprocal dispersion of 1.45 \AA \ pixel$^{-1}$ (we used a 120 $\mu$m-wide slit), thus covering the spectral region 5810--7205 \AA. The slit of the spectrograph was rotated for HD 37742 in order to avoid contamination from the visual companion. The speckle
interferometric observations of Mason \etal (1998) exclude the existence in HD 24912, HD 30614 and HD 151804 of bright ($\Delta m$ $\la$ 3 mag), close companions (0.035$''$ $\la$ $\rho$ $\la$ 1.5$''$). Apart from HD 37742, a search in the {\it Hipparcos} database reveals only one star in our sample with a close visual component (HD 42087: $\Delta H_p$=2.58 mag and $\rho$=0.6$''$). 

The {\tt IRAF}\footnote{{\tt IRAF} is distributed by the National Optical Astronomy
Observatories, operated by the Association of Universities for Research in Astronomy, Inc., under cooperative
agreement with the National Science Foundation.} tasks were used to carry out standard reduction
procedures (i.e., bias subtraction, flat-fielding, extraction of the spectra, and wavelength calibration). Spectra of FeNe lamps were  taken immediately before and/or after the stellar
exposure. The spectra were subsequently put in the heliocentric velocity frame.  The \has line profiles were found
to be significantly affected by telluric lines, as the observations were often carried out under unfavourable
atmospheric conditions (e.g., high humidity). Therefore, we have obtained spectra of the rapid rotator
HD 65810 (A1 V)  to create a library of telluric lines that would be used on the target stars.
Since this procedure proved unsatisfactory, we used instead the high-resolution telluric spectrum of
Hinkle \etal (2000) degraded to our resolution. The water vapour lines in the pseudo-continuum regions around
\has were interactively shifted and scaled with the {\tt IRAF} task {\tt telluric} until the residuals between the
observed spectra and the template telluric spectrum were minimized. The spectra were finally continuum-rectified
by fitting a low-order spline3 polynomial to fixed line-free spectral regions on both sides of H${\alpha}$. We generally obtained three consecutive
exposures to allow a more robust rectification of the spectra and filtering of cosmic ray events. Any individual
exposure with evidence for an imperfect continuum normalization was excluded during the co-adding operation. The
typical signal-to-noise ratio (S/N) at \has in the combined spectrum is quoted for each star in Table 3, and
lies in the range 125--335 per pixel in the continuum.  Large velocity
variations of the photospheric \hes line, which might arise from binary motion, were found in HD 13854 and HD 47240 ($\sigma$$\sim$15 and 30 km s$^{-1}$, respectively). The spectra
were hence realigned and put in the same reference frame. The remaining stars may have companions
(e.g., HD 37742), but the radial velocity variations are small, i.e., comparable to the typical uncertainties in the
wavelength calibration ($\sim$5 km s$^{-1}$). Hence, all the discussed spectral variations should
be intrinsic to the star.

\subsection{Photometry}
Coordinated {\it B}- and {\it I}-band (Johnson) photometric observations were carried out with the 24-inch
telescope of the Bell Observatory operated by the Western Kentucky University (USA). Six stars were observed during 7 nights in January, February
and April 2002: HD 13854, HD 14134, HD 14818, HD 42087, HD 47240, and HD 52382.
A typical observation (repeated up to 3 times per night) comprises up to 10 short-exposure (1--10 sec) CCD images of the target and the
surrounding field (0.53$''$ pixel$^{-1}$ scale, providing an $4.4'\times 6.6'$ field).

The standard  processing routines were performed with {\tt IRAF} tasks and included bias and flat field correction,
followed by measurements of all available comparison stars
in the fields surrounding the targets. Pre-selection of appropriate comparison stars was based on their brightness
and non-variability. All the differential measurements performed on the stars from individual frames were combined
into normal observation, thus producing 1--3 data points per night for each star. This resulted in overall accuracy $\sim$0.01 mag per data point for each comparison star. This relatively low accuracy was mainly dictated by the lack of appropriately bright comparison
stars, the problem being especially acute for HD 14818, HD 42087, and HD 47240. For HD 42087, in particular, there was only one very faint comparison star in the field, resulting in a much lower accuracy ($\sim$0.02 mag). These data are not discussed in the following.

\section{Overview of the survey and quantitative analysis}
The \has time-series are shown in Fig.1, along with the temporal variance spectra (TVS; Fullerton \etal 1996). As can be seen, {\it all} stars display significant variability across most of the \has profile (the extent of the variability in velocity space is well correlated with the wind terminal velocity: $\Delta v$ $\sim$ $\pm$0.3 $v_{\infty}$). One can note that morphological evidence for a \has feature partly formed in the outflow (i.e., a line profile clearly departing from a pure photospheric one) is systematically accompanied by much more prominent changes. The detected variability is qualitatively similar to the cases reported by
Ebbets (1982). Noteworthy is the emission-like episode experienced by HD 14134 and HD 43384. The dramatic variations previously reported in HD 21291 (Denizman \&
Hack 1988; Zeinalov \& Rzaev 1990) and HD 91316 (Smith \& Ebbets 1981) are not seen in our data. 

Most changes take place
on a daily time-scale, although significant hourly variations are sometimes observed. These subtle, short-term
changes should be considered with some caution. In some cases,
we cannot exclude the possiblity that they largely arise from an imperfect continuum
normalization and/or removal of telluric features (Section 3.1). To better assess the reality of these short-term
variations, we use a 'quality flag' defined as the number of co-added consecutive exposures (indicated for each
spectrum in Fig.1). Hourly changes between spectra rated as '1' should be regarded with some suspicion
(e.g., HD 14134), whereas they are very likely to be real between spectra rated as '3'
(e.g., HD 37128). Our data support the hourly changes found in HD 38771 by Rusconi \etal (1980), but not the variations on a time-scale of 30 minutes reported for HD 53138 by van Helden (1972). 

\begin{figure*}
\epsfbox{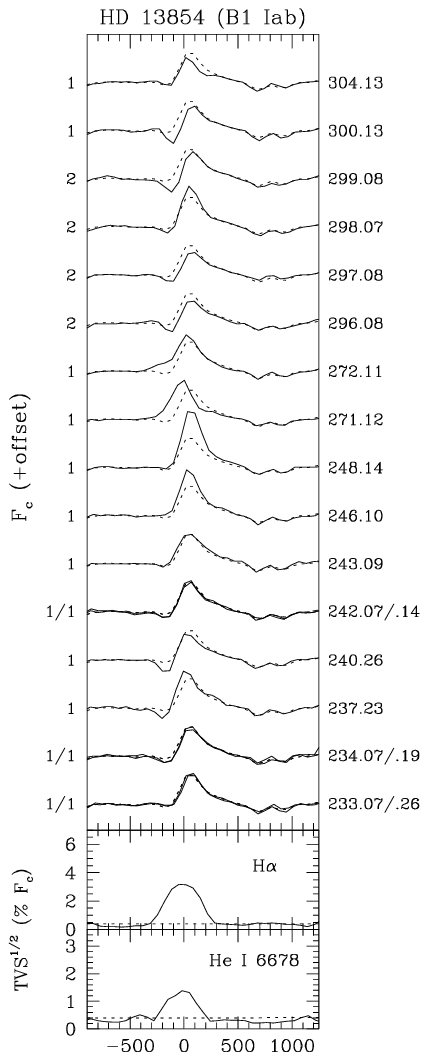}
\epsfbox{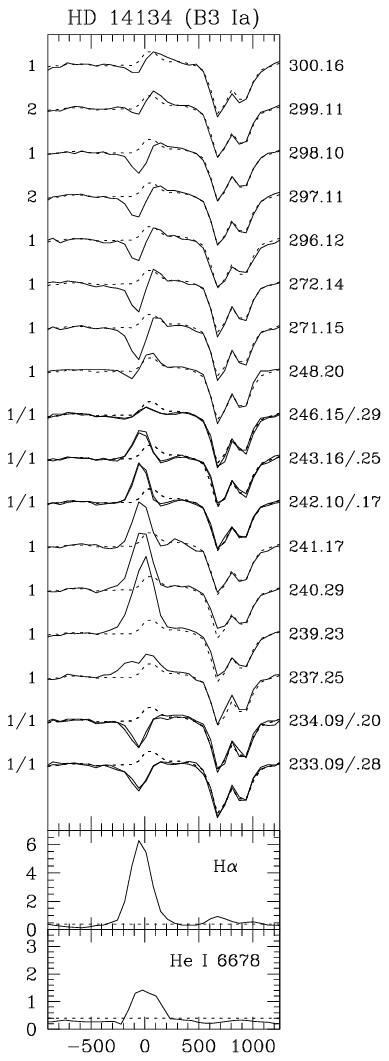}
\epsfbox{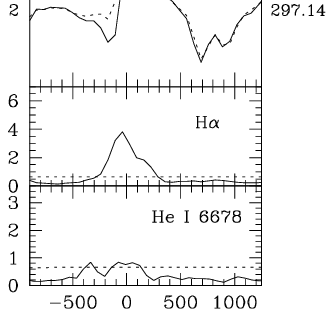}
\epsfbox{fig1d.ps}
\epsfbox{fig1e.ps}
\epsfbox{fig1f.ps}
\epsfbox{fig1g.ps}
\epsfbox{fig1h.ps}
\vspace*{-64.2cm}
\caption{\has time-series for HD 13854, HD 14134, HD 14818, HD 21291, HD 24398, HD 24912, HD 30614, and HD 31327.
The spectra are displayed in the stellar rest frame. Consecutive spectra are shifted by 0.15, 0.10, 0.10, 0.07, 0.10, 0.10, 0.10, and 0.10 continuum units, respectively. The mean profile is overplotted as a dashed line. The numbers
to the left-hand and right-hand sides of the upper panels give the number of consecutive exposures used to
form the corresponding spectrum and the mean Julian date of the observations ($HJD$--2,452,000),
respectively. The two bottom portions of each panel show the TVS of \has and \hes (Fullerton \etal 1996), along with the threshold for a significant
variability at the 99.0 per cent confidence level ({\it dashed line}).}
\end{figure*}

\addtocounter{figure}{-1}
\begin{figure*}
\epsfbox{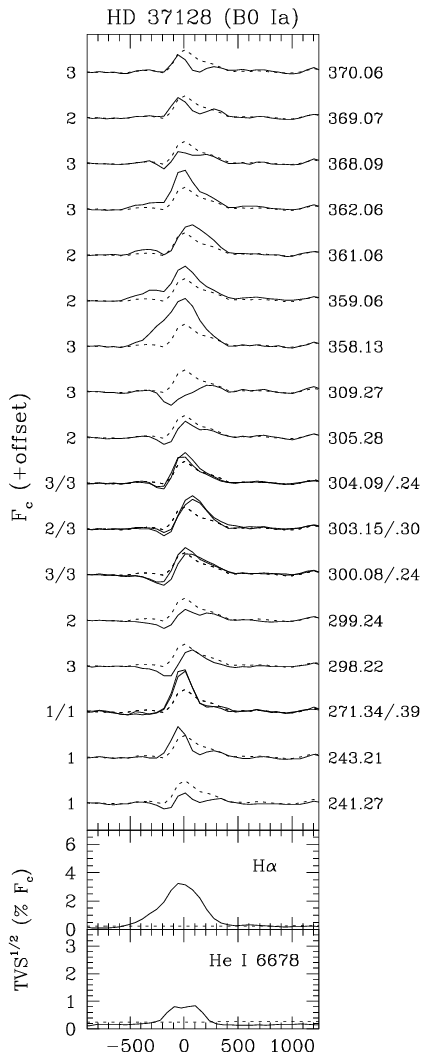}
\epsfbox{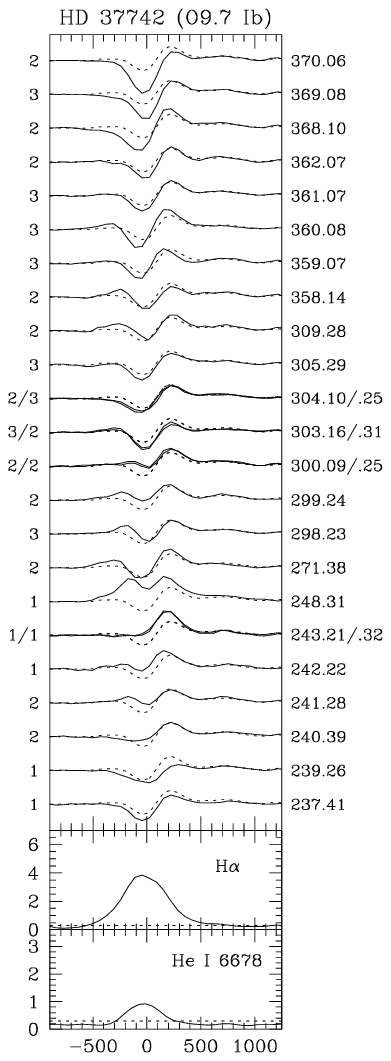}
\epsfbox{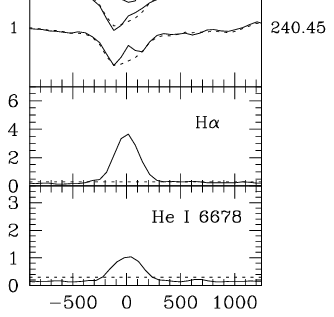}
\epsfbox{fig1l.ps}
\epsfbox{fig1m.ps}
\epsfbox{fig1n.ps}
\epsfbox{fig1o.ps}
\epsfbox{fig1p.ps}
\vspace*{-64.0cm}
\caption{\has time-series for HD 37128, HD 37742, HD 38771, HD 41117, HD 42087, HD 43384, HD 47240, and HD 52382. Consecutive spectra are shifted by 0.12, 0.13, 0.08, 0.20, 0.10, 0.12, 0.15, and 0.10 continuum units, respectively.}
\end{figure*}

\addtocounter{figure}{-1}
\begin{figure*}
\epsfbox{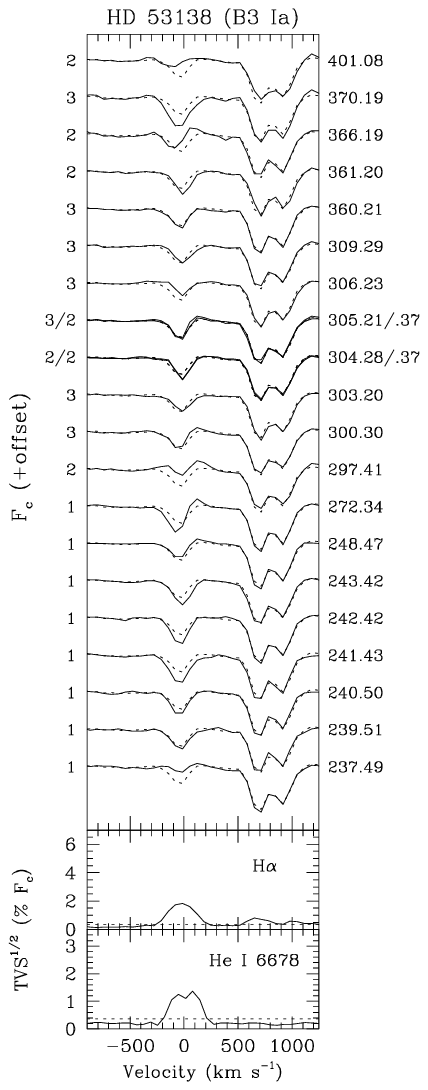}
\epsfbox{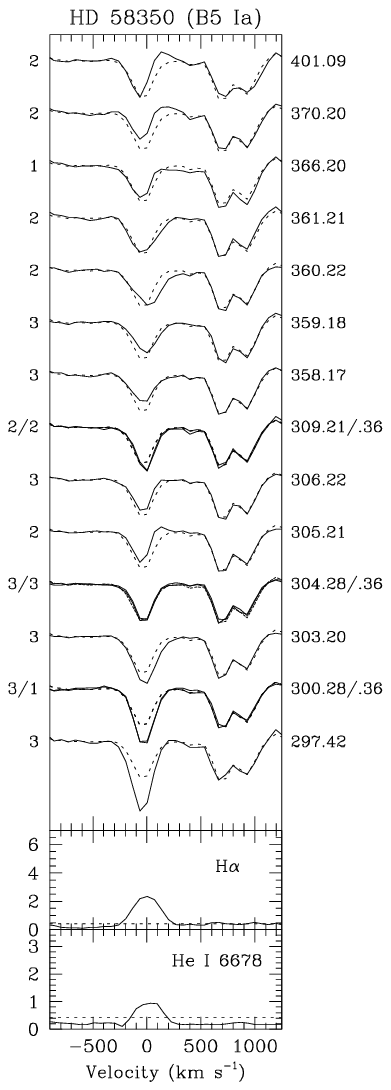}
\epsfbox{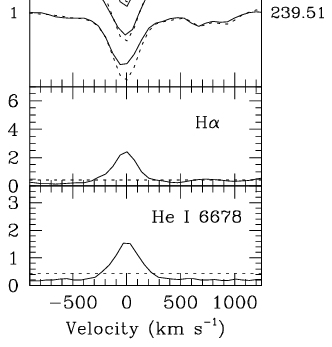}
\epsfbox{fig1t.ps}
\vspace*{-32.0cm}
\caption{\has time-series for HD 53138, HD 58350, HD 91316, and HD 119608. Consecutive spectra are shifted by 0.10 continuum units in all cases.}
\end{figure*}

One should bear in mind that the observed variations in \has may not be straightforwardly associated to wind activity. It is likely that time-dependent changes, possibly arising from pulsations, as commonly observed in early-type supergiants (e.g., de Jong \etal 1999; Kaufer \etal 1997), also affect the underlying photospheric profile. Although our modest spectral resolution precludes a detailed study, we choose to illustrate the temporal changes in photospheric lines by means of \he.\footnote{We do not discuss the variations that may affect the potentially interesting He I $\lambda$5876 line because of difficulties in defining the neighbouring continuum, as well as edge effects.} With the exception of  HD 152236 which occasionally exhibits a P Cygni profile, this feature appears to be primarily of photospheric origin for all the stars in our sample. The changes affecting this line are generally significant (see Fig.1), and are likely paralleled in the \has photospheric component. Because of substantial pressure broadening in B-type supergiants, such variations will extend in velocity space well beyond the projected rotational velocity (e.g., Ryans \etal 2002). In many cases the observed \hes variability barely exceeds the detection thresholds, which prevents us from linking the patterns observed in \has and \he.  The limited spectral resolution also precludes a straightforward interpretation of the variability as arising either from a variable, incipient emission or from temporal fluctuations in the shape of the absorption profile (e.g., HD 31327). A direct comparison between the patterns of variability in \hes and \has will be presented for two illustrative cases in Section 6.1.

\addtocounter{figure}{-1}
\begin{figure}
\epsfbox{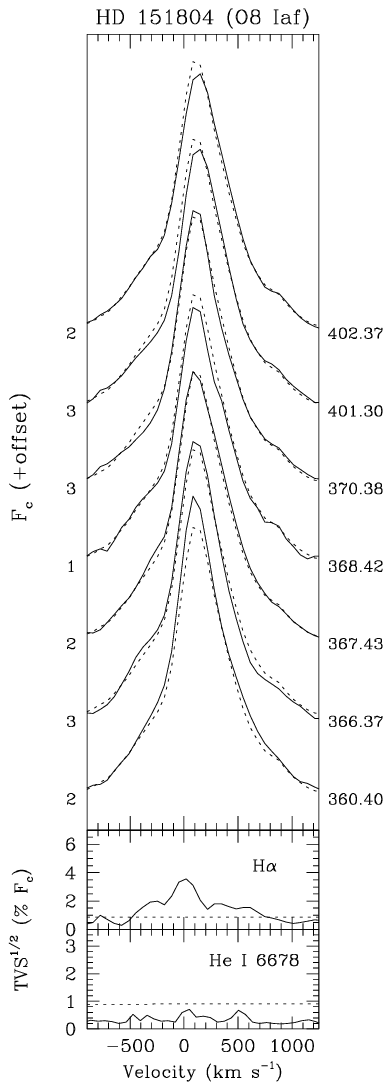}
\epsfbox{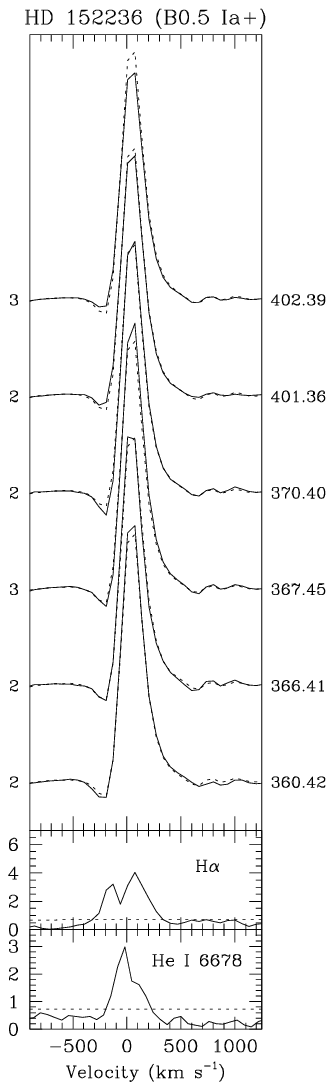}
\vspace*{-10.7cm}
\caption{\has time-series for HD 151804 and HD 152236. Consecutive spectra are shifted by 0.20 and 0.40 continuum units, respectively.}
\end{figure}

To quantify the level of spectral and photometric variability, we define the two activity indices $a_{\rm phot}$ and $a_{\rm lpv}$ (Table 4). The former was calculated both from {\it Hipparcos} data
and from our photometric observations, and is defined as: $a^2_{\rm phot}$=$\sigma^2_{\rm obs}$--$\sigma^2_{\rm instr}$,
where $\sigma_{\rm obs}$ and  $\sigma_{\rm instr}$ are the scatter of the observations and the instrumental
noise, respectively (see Marchenko et al. 1998). The spectral variability index, $a_{\rm lpv}$, is the fractional
amplitude of the line-profile variations, i.e., the amplitude of the changes normalized by the strength
of the feature (see equation [15] of Fullerton \etal 1996). This index was calculated both for \has and for \he.

\begin{table}
\centering
\caption{Photometric and spectroscopic activity indices.}
\hspace*{-1.0cm}
\begin{tabular}{lccccc}
\hline
Name & \multicolumn{2}{c}{$a_{\rm phot}$} & \multicolumn{2}{c}{$a_{\rm lpv}$} \\
     &  This study & {\it Hipparcos} & \has & \hes\\
     & (mmag) & (mmag) & (per cent) & (per cent)\\\hline
HD 13854                  & 26 & 13 & 47.3 & 10.6\\
HD 14134 (V520 Per)       & 22 & 24 & 259  & 11.2\\
HD 14818 (10 Per)         & 20 & 17 & 34.8 & 6.01 \\
HD 21291                  &    & 17 & 12.5 & 0$^{b}$\\
HD 24398 ($\zeta$ Per)    &    & 0  & 5.91 & 5.57\\
HD 24912 ($\xi$ Per)      &    & 0  & 14.6 & 6.53\\
HD 30614 ($\alpha$ Cam)   &    & 9  & 17.8 & 10.7\\
HD 31327                  &    & 0  & 6.45 & 7.92\\
HD 37128 ($\epsilon$ Ori) &    & 11 & 81.9 & 7.66\\
HD 37742 ($\zeta$  Ori A) &    & 4  & 82.7 & 11.0\\
HD 38771 ($\kappa$ Ori)   &    & 9  & 32.6 & 7.35\\
HD 41117 ($\chi^2$ Ori)   &    & 15 & 27.9 & 12.8\\
HD 42087 (3 Gem)          & 20: & 13 & 91.2 & 8.98\\
HD 43384 (9 Gem)          &    & 20 & 135  & 7.69\\
HD 47240                  & 17 & 5  & 40.2 & 5.89\\
HD 52382                  & 20 & 16 & 32.1 & 11.0\\
HD 53138 ($o^2$ CMa)      &    & 21 & 66.4 & 9.64\\
HD 58350 ($\eta$ CMa)     &    & 28 & 44.1 & 6.90\\
HD 91316 ($\rho$ Leo)     &    & 9  & 12.8 & 10.1\\
HD 119608                 &    & 0  & 10.2 & 0$^{b}$\\
HD 151804          &    & 0$^{a}$  & 5.16 & 0$^{b}$\\
HD 152236 ($\zeta^1$ Sco) &    & 22 & 5.66 & 19.1\\\hline
\end{tabular}\\
\begin{flushleft}
$^{a}$ Photometric variable according to Balona (1992), and references therein.\\
$^{b}$ These stars do not exhibit significant variability in \hes (see Fig.1).
\end{flushleft}
\end{table}

\section{Period search}
We performed a period search analysis by calculating the power spectra (hereafter PS) using the
technique of Scargle (1982) on: (a) the photometric {\it Hipparcos} data (our observations are not
amenable to a period search), (b) the pixel-to-pixel \has time-series, (c) the equivalent widths (EWs)
measured within fixed wavelength intervals encompassing the whole of \has ({\it even} in the case of P Cygni profiles).
Despite the fact that the EW and  pixel-to-pixel line-profile variations are not, strictly speaking, two independent quantities, a period search in the two data sets helps assessing the reality of the detected signals. Subsequent correction of the frequency spectrum by the CLEAN algorithm was performed in order to
remove aliases and spurious features induced by the uneven spacing  of the data in the time
domain (Roberts, Leh\'ar \& Dreher 1987).

\subsection{Photometric data set}
Periodicities in the range 1--27 days have been found for 9 stars in our sample. Several stars display significant variability at the 0.03--0.07 mag level without, however,
evidence for cyclical changes. A complex, multi-periodic behaviour cannot be ruled out in this case. As can be seen in Fig.2, our photometric observations of HD 13854,
HD 14134, HD 14818, and HD 47240 support the periodic nature of the variations suggested by the {\it Hipparcos} data.
This demonstrates the remarkable long-term (years) coherency in the patterns of variability: folded
with the appropriate periods, practically all the
new data points stay well within the scatter limits (mainly intrinsic, considering the typical accuracy of the
{\it Hipparcos} photometry), {\it without} any adjustment of the initially chosen zero phases.

\begin{figure}
\epsfbox{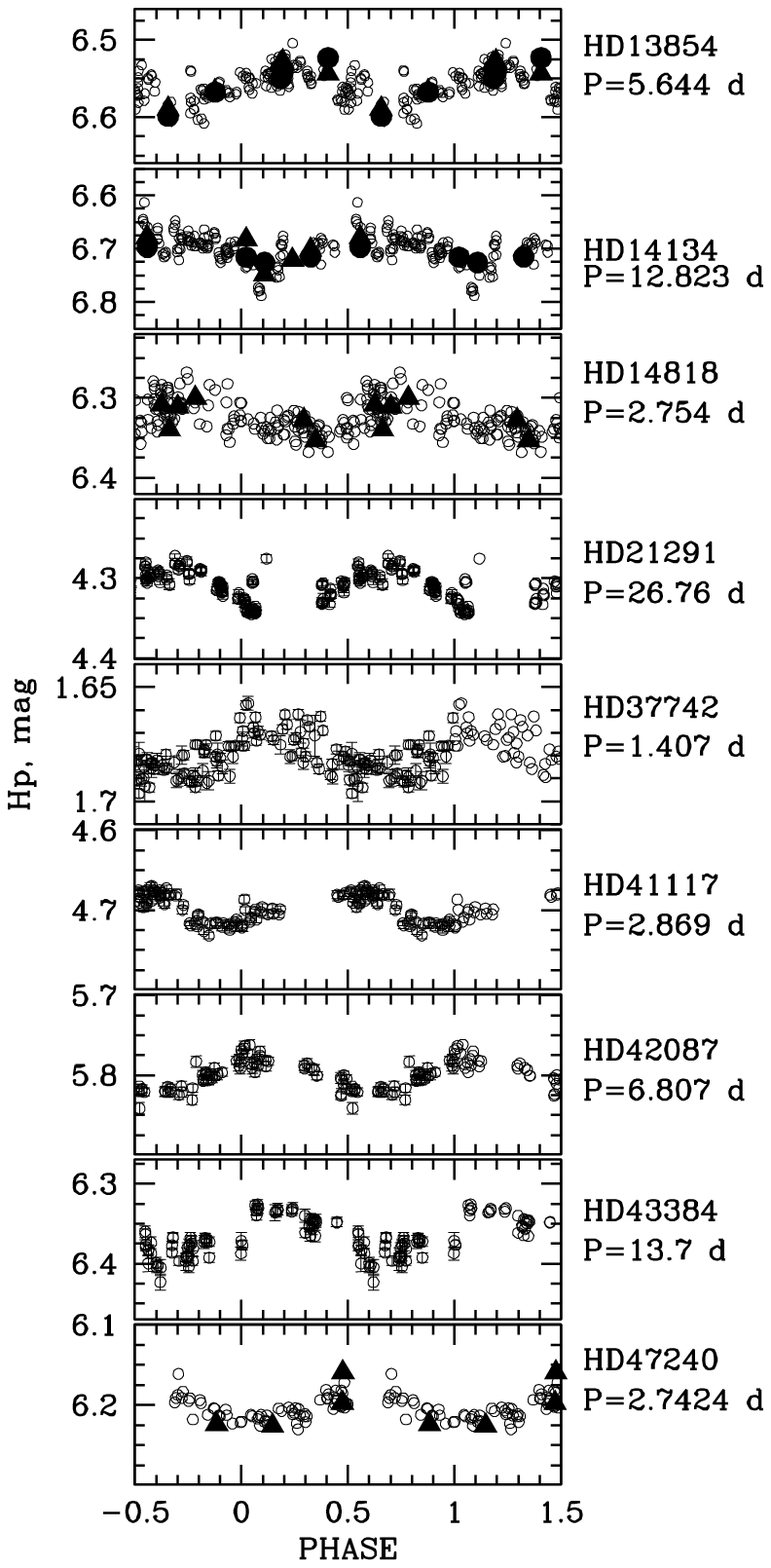}
\caption{{\it Hipparcos} light curves for stars displaying periodic variations. Zero phase is arbitrarily assigned to the
start of the observations. The size of the
symbols approximately corresponds to 2-$\sigma$ error bars (if not shown). Our {\it B}- and {\it I}-band
observations are overplotted as filled triangles and circles, respectively (the points are vertically shifted by a constant value to match the {\it Hipparcos} data).  In this case the size of the symbols corresponds
to 4-$\sigma$ error bars. }
\end{figure}

\subsection{Spectroscopic data set}
The significance of the peaks in the PS can be, in principle, rigourously estimated after normalization by
the variance of the data (e.g., Scargle 1982). In the case of unevenly spaced time-series, however, the false
alarm probability (hereafter FAP) is only indicative and cannot be used as a robust indicator of the presence
of periodic signals in the data (see, e.g., Antokhin \etal 1995). While assessing the reality of the periods
found in our spectroscopic data set, we therefore took into account additional criteria. Namely, we inspected the
period-folded EW/line-profile data, and compared the EW and 2D pixel-to-pixel PS
(even though the line-profile variations are obviously not necessarily accompanied by significant EW changes).
The existence of a
cyclical pattern of variability is suggested for 9  stars out of the 17 included in the analysis. These cases
are individually discussed below. No indication of a cyclical behaviour was found in HD 24398, HD 30614,
HD 31327, HD 43384, HD 52382, HD 53138, HD 58350, and HD 91316. Kaper \etal (1998) and Markova
(2002) report variations on a time-scale of about one week in HD 30614. A search for periodic modulations
of the UV lines in HD 53138 proved inconclusive, in line with our non detection (Prinja \etal 2002).\\
\\
{\bf HD 13854:} Significant power at $\nu$$\sim$0.955 d$^{-1}$ is found in the EW data (Fig.3),
but not in the \has time-series. \\

\begin{figure}
\epsfbox{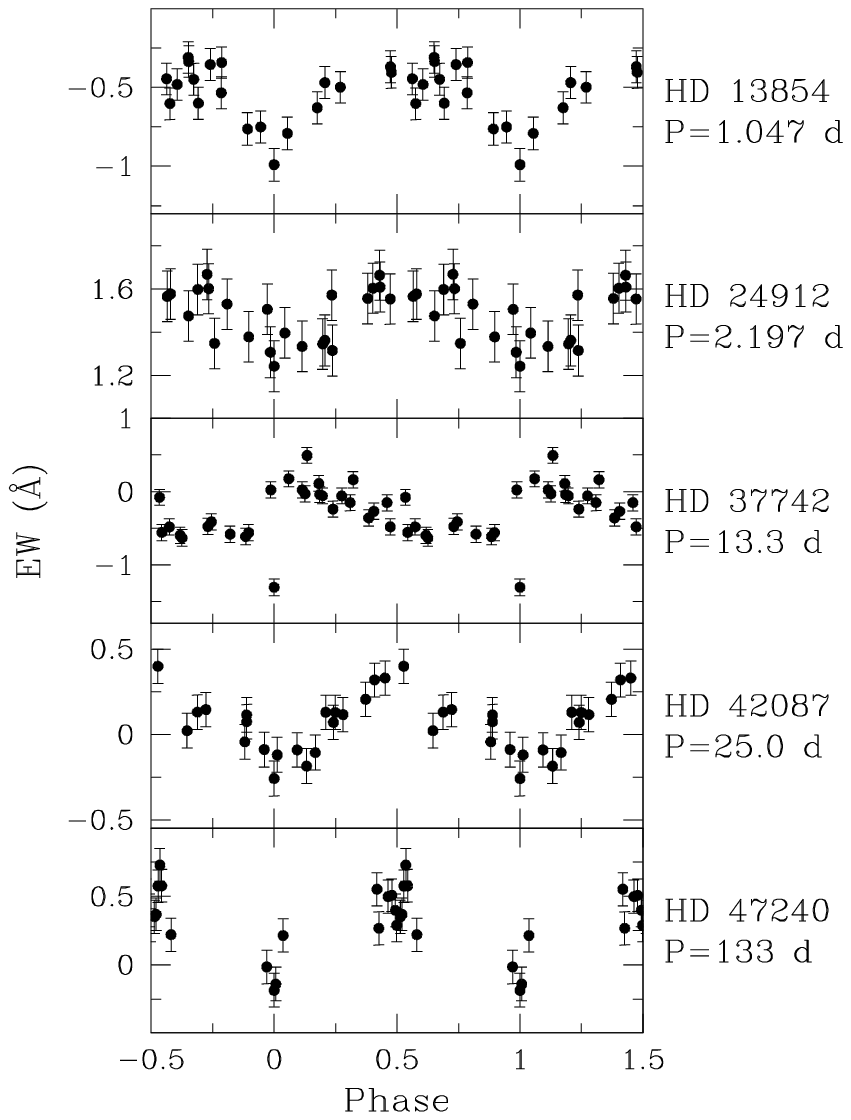}
\caption{EW of the \has line as a function of phase (data for HD 14134 in Fig.5). Zero phase is arbitrarily set at maximum emission (or, alternatively, at minimum absorption). The uncertainties have been calculated following Chalabaev \& Maillard (1983).}
\end{figure}

\hspace*{-0.7cm} {\bf HD 14134:} This star does not exhibit any evidence for binarity (Abt \& Levy 1973).
The photometric variability of HD 14134 is relatively well documented (Rufener \& Bartholdi 1982;
Waelkens \etal 1990; Krzesi\'nski, Pigulski \& Ko\l aczkowski 1999; Adelman, Y\"uce \& Engin 2000), but the {\it Hipparcos} data provide the first clear evidence for a periodic pattern with $\cal P$$\sim$12.823 d (Fig.2). Two periodic signals at $\nu_1$=0.045 and $\nu_2$=0.080 d$^{-1}$ (${\cal P}$$\sim$22.2 and 12.5 d,
respectively) are unambiguously detected in the 2D PS, with FAP$\la$0.1 per cent (see Fig.4). The latter frequency is also found in the EW PS (once again with FAP$\la$0.1 per cent) and is, within the errors, identical to the value found in the photometric data. The non-detection of $\nu_1$ both in the EW and in the photometric data sets leads us to take $\nu_2$ as
the fundamental frequency. The two signals are likely to be simple harmonics ($\nu_2$$\sim$2$\nu_1$).
 As can be seen in Fig.5, maximum \has emission nearly coincides with maximum light. We use the photometric ephemeris when folding the EWs ($\cal{P}$=12.823 days and $T_0$=2,447,867.8), as this period is undistinguishable, but more accurate, than the value derived from the spectroscopic data. 

The clearly phase-locked nature of the variations is reminiscent
of some early-type stars displaying evidence for a decentered, dipole magnetic field (e.g., Stahl \etal 1996). A single circular polarization spectrum of HD 14134 was therefore obtained on 29
July 2003 using the MuSiCoS spectropolarimeter at the 2-m Bernard Lyot
telescope (Pic du Midi Observatory). The peak S/N was 250 (weather
conditions were relatively poor). The mean profiles obtained using
the Least-Squares Deconvolution procedure (LSD; Donati \etal 1997) show no
significant circular polarization within spectral lines. The measured
longitudinal magnetic field is 45$\pm$145 G (G. A. Wade, private communication). \\
\\
{\bf HD 24912:} de Jong \etal (1999) discussed periodic changes in photospheric lines that may
arise from pulsations ($\cal{P}$$\sim$3.45 hours). This star presents a strong case for a persistent, cyclical
variability in the UV wind line-profiles with ${\cal P}$$\sim$2 days (Kaper \etal 1997). From an analysis of an extensive data set of UV and optical observations, de Jong \etal (2001)
reported a 2.086-d modulation of the line profiles, notably in \ha. In our data we find indication for a period tantalizingly close to,
but apparently significantly different from this value: ${\cal P}$$\sim$2.20 d (with a typical uncertainty
of 0.03 d). There is also some evidence for long-term changes in the pattern of variability. While cyclical variations are suggested here in the red wing of \has (Fig.4), such changes predominantly occured in the blue wing on October 1994 (de Jong \etal 2001). Although the periodic signal found in our data is present both in the \has line-profile and EW time-series,
we warn that its formal significance is low in the 2D PS (Fig.4), while the phase-locked nature of the
EW variations is only marginal (Fig.3).  \\
\\
{\bf HD 37128:} Prinja \etal (2002) discussed the UV line-profile variations in this star, but the
limited time span of the {\it IUE} observations ($\sim$17 hours) precluded the detection of
a cyclical behaviour on a rotational time-scale. Three signals are detected at the 99.0
per cent confidence level in the pixel-to-pixel \has PS at: $\nu_1$=0.055, $\nu_2$=1.070,
and $\nu_3$=1.280 d$^{-1}$ (Fig.4). $\nu_2$ can be identified as the 1-d alias of $\nu_1$.
Interestingly, the highest peaks in the EW PS correspond to the two remaining signals, although
with low significance level, as confirmed by the poorly coherent patterns of variability. Our
data therefore suggest the existence of two periods at ${\cal P}$$\sim$18.2 and 0.78 d, but at a low confidence level. \\
\\
{\bf HD 37742:} Line-profile variations on a time-scale of about 6 days as suggested both in \has and in UV wind lines (Kaper \etal 1998, 1999). Several significant peaks (all with FAP$\la$1 per cent) are found in the pixel-to-pixel PS at:
$\nu_1$=5$\times$10$^{-3}$, $\nu_2$=0.020, $\nu_3$=0.880, and $\nu_4$=1.148 d$^{-1}$
(Fig.4). Only $\nu_4$ is found in the EW data set, but this signal is likely spurious, as judged from the lack of coherent variations. Alternatively, the EW variations appear reasonably phase-locked when folded with $\nu_5$=0.075 d$^{-1}$ (Fig.3). This signal is also found in the pixel-to-pixel PS, but the confidence level is low. \\
\\
{\bf HD 38771:} Two signals are found in the 2D PS (see Fig.4): at $\nu_1$=0.210 (FAP$\sim$5 per cent)
and $\nu_2$=0.955 d$^{-1}$ (FAP$\sim$1 per cent), but there is no evidence for periodicity in the EW
time-series.\\

\begin{figure*}
\epsfbox{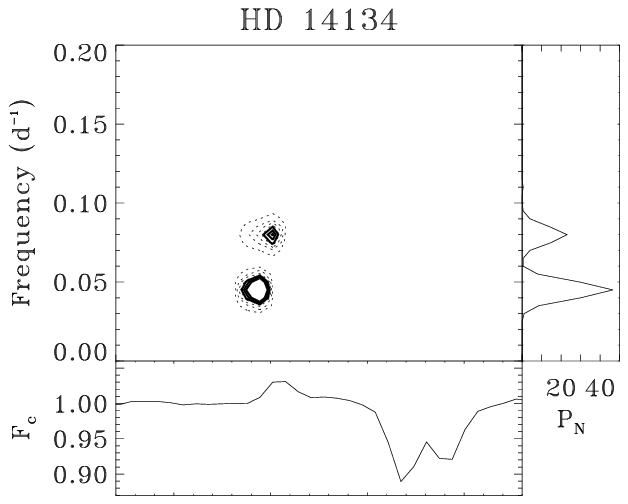}
\epsfbox{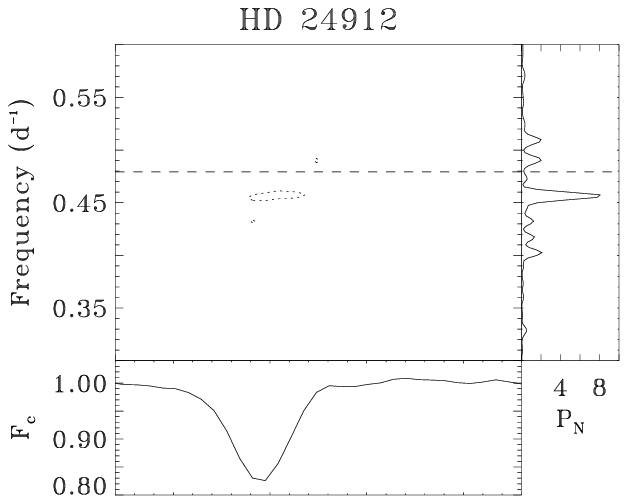}
\epsfbox{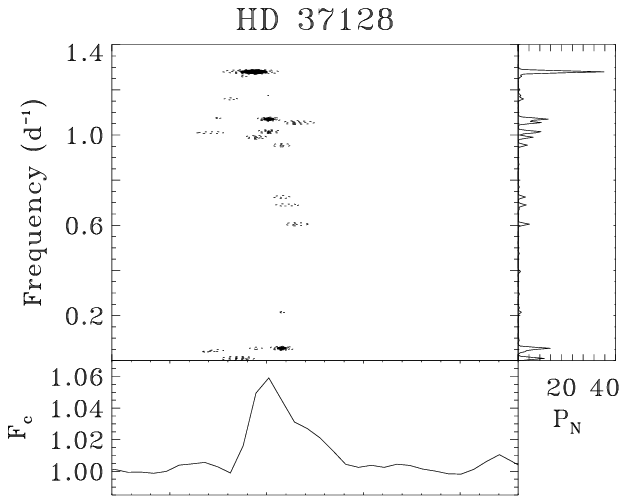}
\epsfbox{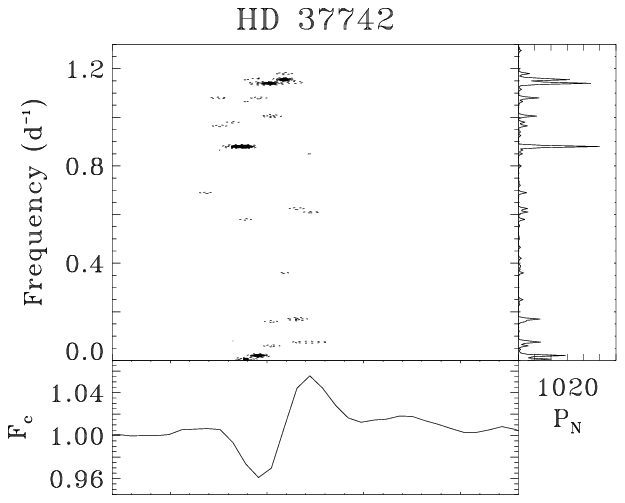}
\epsfbox{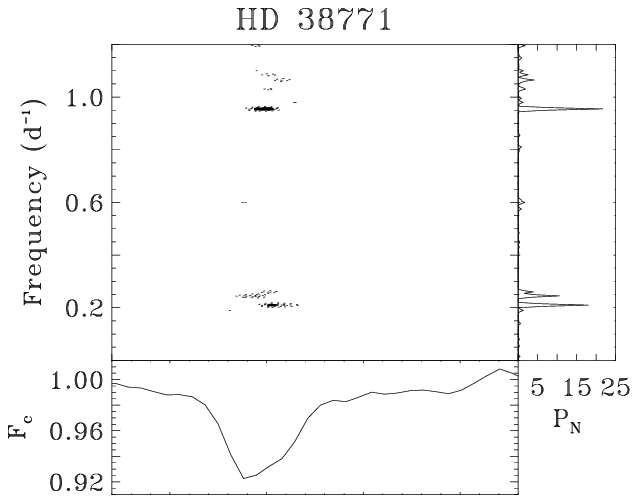}
\epsfbox{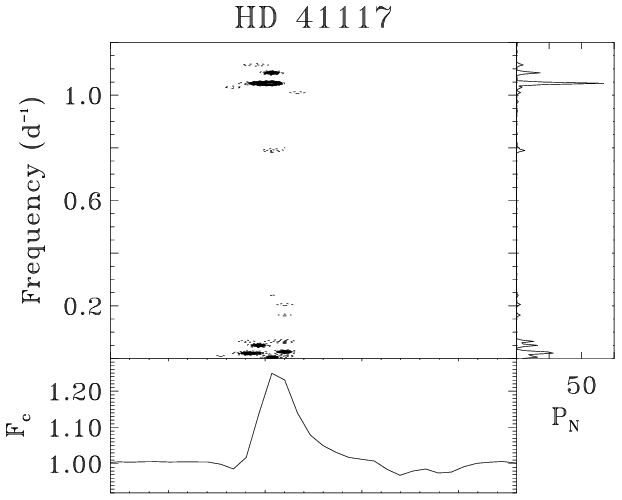}
\epsfbox{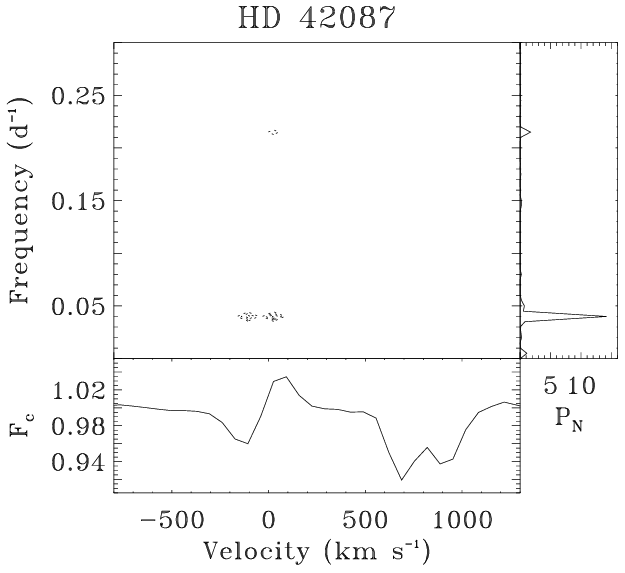}
\epsfbox{fig4h.ps}
\vspace*{-22.3cm}
\caption{Pixel-to-pixel power spectra of the \has time-series. The number of iterations and gain of the CLEAN algorithm were typically set to 2000 and 0.6, respectively. All the PS are variance-normalized to provide a uniform assessment of significance. The solid contours are drawn for intensities of 12.9, 10.6, and 8.0 (this corresponds to significance levels of 99.9, 99.0, and 95.0 per cent, respectively). The dotted contours are drawn for intensities of 7.0, 5.0, 3.0, and 1.0. In each diagram the frequency range is chosen to show more clearly the detected signals; no significant power peaks are found outside this frequency domain. For HD 24912, the dashed line indicates the 2.086-d period found by de Jong \etal (2001). The bottom section of each panel shows the mean spectrum (in the stellar rest frame), while the side panel presents the PS summed over the shown velocity range.}
\end{figure*}

\begin{figure*}
\epsfbox{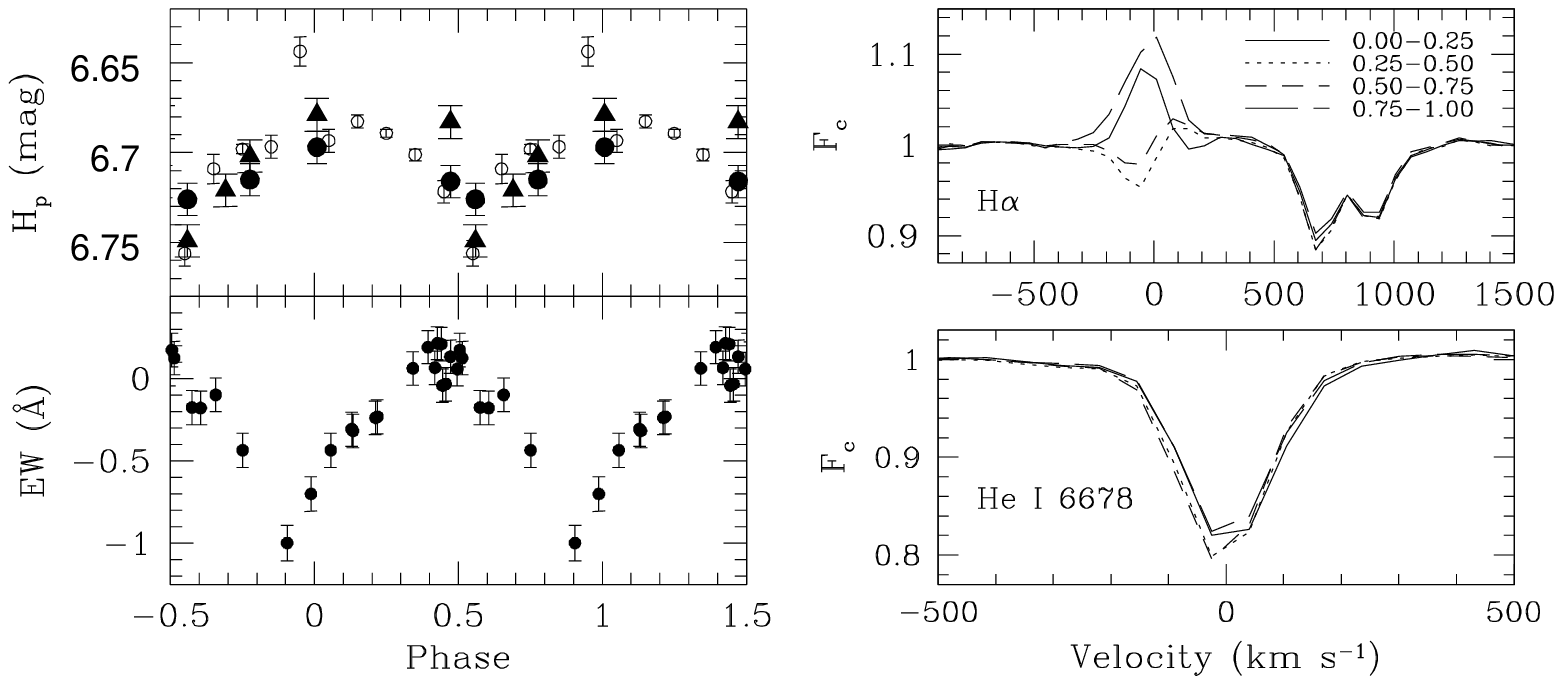}
\caption{{\it Left-hand panels}: light curve of HD 14134 binned to 0.1-phase resolution (symbols as in Fig.2. Here note that zero phase is arbitrarily set at maximum light) and EWs of the \has line, as a function of phase. {\it Right-hand panels}: phase-averaged \has and \hes profiles. In all cases we use the photometric ephemeris: $\cal{P}$=12.823 days and $T_0$=2,447,867.8.}
\end{figure*}

\hspace{-0.7cm} {\bf HD 41117:} Several peaks are found in the 2D PS (all with FAP$\la$1 per cent) at:
$\nu_1$=5$\times$10$^{-3}$, $\nu_2$=0.025 (and its first harmonic at 0.050 d$^{-1}$),
$\nu_3$=1.045, and $\nu_4$=1.085 d$^{-1}$ (Fig.4). The highest peaks in the EW PS are found at
$\nu_5$=0.025, $\nu_6$=0.065, and $\nu_7$=0.205 d$^{-1}$. Among these, $\nu_6$ and $\nu_7$ can
be ruled out from the lack of phase-locked EW variations. Despite the detection of $\nu_5$ both in the 2D and in the EW PS, the poorly coherent pattern of variability casts serious doubt on the
existence of this 40-d signal. \\
\\
{\bf HD 42087:} This star does not exhibit significant radial velocity variations (Mayer \etal 1999). The highest peak in the 2D PS is found at $\nu_1$=0.040 d$^{-1}$, and appears both in the emission
component and in the absorption trough of the \has P Cygni profile (Fig.4). Although its formal
significance is relatively low (FAP$\sim$5 per cent), the existence of a 25-d periodicity is supported by the EW data (Fig.3). \\
\\
{\bf HD 47240:} There is a tentative indication that the changes in Si\,IV $\lambda$1400 are repetitive
over a 3- to 5-d time-scale (Prinja \etal 2002). An ill-defined, long-term periodic signal is found in the EW
data set at $\nu_1$=0.0075 d$^{-1}$, along with smaller peaks at $\nu_2$=0.105 and $\nu_3$=0.247 d$^{-1}$.
The two latter frequencies are likely spurious, as judged from the poorly coherent nature of the phased
EW variations. Interestingly, $\nu_1$ is also found in the pixel-to-pixel \has PS with a confidence level
close to 95 per cent (Fig.4). However, both the close correspondence with the total time span of the
observations (see Table 3) and the sparsely sampled EW data (Fig.3) lead us to discard it from further
consideration. Furthermore, the \has morphology is typical of a fast-rotating star seen nearly equator-on
(Conti \& Leep 1974; Petrenz \& Puls 1996). It is therefore unlikely that spectral changes operating on
such a long time-scale would be associated to rotation. Variations intrinsic to a circumstellar disc-like
enhancement cannot be ruled out a priori, but they would manifest themselves as V/R variations of
the line emission peaks, which are not observed (Fig.4).\\

\section{Discussion}
\subsection{Incidence of cyclical line-profile variations}
The results of our period search are summarized in Table 2. Several of the listed periods must be regarded
as tentative, especially those detected either in the 2D or in the EW PS. Independent confirmation is clearly
needed in this case. We also recall that HD 13854 and HD 47240 might be part of a close, massive binary system (Section 3.1). Since our survey was not designed to detect short-term periodicities, the found periods close to one day must also be regarded with some caution. They might arise from rotational modulation
only for the fastest rotators. Conversely, they might be associated with the pseudo-periodic ejection of 'puffs'
of material. Such an outward propagation of density perturbations has been proposed to account for the recurrent
changes (with a time-scale of $\sim$20 days) in the \has line of the B0.5 hypergiant HD 152236
(Rivinius \etal 1997). Changes on a time-scale of about one day might be compatible with the characteristic
flow time through the \has line-formation region in our program stars, provided that these putative,
spatially-localized density enhancements follow a $\beta$-type velocity law with an exponent close to
unity.\footnote{Here we do not consider the outward propagation of small-scale clumps of material resulting from the unstable nature of radiatively-driven winds, as this would lead to a {\it stochastic} pattern of variability (e.g., Eversberg, L\'epine \& Moffat 1998).}
Although such an interpretation cannot be completely ruled out, it should be noted that HD 152236 is
considered as a candidate luminous blue variable and is likely not representative of the stars in our sample
(Sterken \etal 1997). Pulsations can also operate on a 1-d time-scale, but we find no trivial relation between the
spectroscopic and photometric periods (the latter presumably related to pulsations, as argued below).

Summarising, we consider that the two stars with a relatively
long-term periodic modulation seen both in the EW and in the line-profile data sets present the strongest cases for a cyclical behaviour: HD 14134 and HD 42087 (with ${\cal P}$$\sim$12.8 and 25.0 d, respectively). The phase-related patterns of variability are remarkably similar (see Fig.6). In addition, the line-profile changes in \has and \hes also seem to be correlated in both stars, with excess emission in \has being accompanied by a weakening of \he. As can be seen in Fig.5, this may also apply to weak metal lines such as C\,II $\lambda$6578.05 and C\,II $\lambda$6582.88 (note that variations in the continuum level may be invoked in the case of HD 14134 in view of the large, phase-locked photometric variations). Since a significant amount of incipient wind emission is not expected for stars with such relatively low mass-loss rates (Table 1), this phenomenon may point to a link between the wind variability and photospheric disturbances. Higher resolution data are, however, needed to reveal an organized pattern of variability in photospheric lines, as observed in other B-type supergiants (Kaufer \etal 1997, 2002).

The lack of cyclical variations for several stars in our sample can have a physical origin or may be ascribed
to the limitations of the current data set (e.g., insufficient time sampling). Concerning the latter point, it is suggestive
that the stars with a detected period in \has are among those displaying the highest level of variability. The large
amplitude of the variations exhibited by most stars suggests that an imperfect removal of telluric features or S/N issues are probably irrelevant. One may be more concerned by  the fact that our observations were
relatively few and spread out over several months. If we accept the idea that large-scale wind streams give
rise to the variability, substantial variations in the global wind morphology might be expected on a time-scale of months/years (as in the case of the B0.5 Ib star HD 64760; Fullerton \etal 1997). A lack of coherency in the
pattern of variability on a monthly time-scale would thus not be surprising, especially
for the fastest rotators in our sample. It is unlikely in this case that a strong periodic signal would appear in the data.
Another issue which may lead to a loss of periodic signal is the complex pattern of variability presented
by \has in B supergiants (see, e.g., Kaufer \etal 1996;
de Jong \etal 2001; Prinja \etal 2001), emphasizing the need for high spectral/time resolution.

\begin{figure*}
\epsfbox{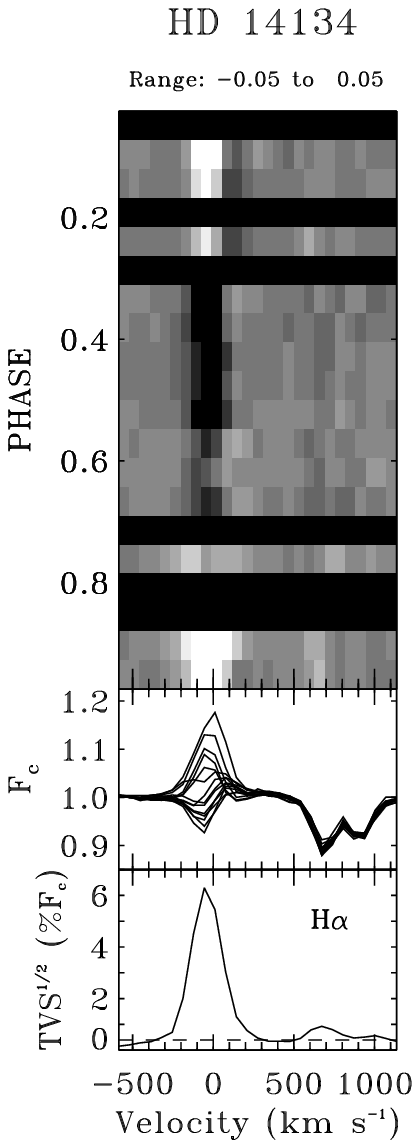}
\epsfbox{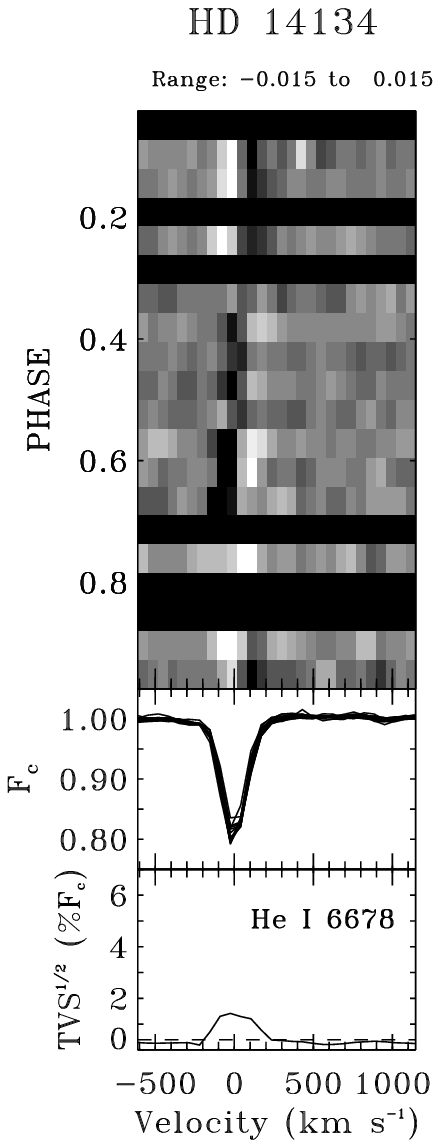}
\epsfbox{fig6c.ps}
\epsfbox{fig6d.ps}
\vspace*{-35.5cm}
\caption{Grey-scale plots of the residuals of \has and \he, as a function of phase, for HD 14134 (${\cal P}$=12.823 d) and HD 42087 (${\cal P}$=25.0 d). The residual profiles are the individual spectra binned to 0.05-phase resolution minus the mean spectrum. Excess emission (or, alternatively, weakening of absorption) appears brighter in these plots. We use the photometric ephemeris in the case of HD 14134, while zero-phase is arbitrarily fixed at maximum emission for HD 42087. The middle portion of each panel shows the superposition of the binned profiles. The TVS, along with the horizontal dashed line indicating the 99.0 per cent confidence level for significant variability, are displayed in the lower portion of each panel. The grey-scale plots are displayed in the stellar rest frame.}
\end{figure*}

\subsection{Evidence for rotational modulation?}
Detailed studies of the line-profile variations in the UV domain have inferred a rather simple geometrical structure for the wind of some B supergiants, with generally two (or four) symmetrical, spatially-extended wind streams (e.g., Fullerton \etal 1997).
Consequently, one generally expects the recurrence time-scale of the line-profile variations to be an integer
fraction of the stellar rotational period, rather than the period itself. In HD 24912, for instance, the period found
in \has can be identified, within the uncertainties, with one half of the rotation period (de Jong \etal 2001). The
fact that the periodicity found in the spectroscopic data set for HD 42087 is longer than the
estimated maximum rotational period is therefore puzzling in this respect (Table 2). Apart from
rotation, however, there is a clear indication for the existence of an additional (perhaps dominating) line
broadening mechanism in B-type supergiants (plausibly tangential macroturbulence;
Howarth 2004). The projected rotational velocities are thus likely to be grossly overestimated for the stars in our sample, with the result
that the true rotational periods may be significantly longer than anticipated
(Ryans \etal 2002). The modulation time-scale of \has in HD 14134 and HD 42087 is commensurate with the
probable rotation periods, but rotational modulation cannot be firmly established in view of these uncertainties. For the former, the 12.8-d period found is close to the estimate of the minimum rotational
period ($\sim$10 days), which would imply that this star is a fast rotator viewed almost equator-on. However,
because in HD 14134  the $v \sin i$ is typical of stars of the same spectral type (Howarth \etal 1997), this 12.8-d time-scale is more likely to be  an integer fraction of the rotation period.

The prominent changes affecting the P Cygni absorption component of the \has line in HD 14134 and HD 42087 (Fig.6) may diagnose variations in the amount of wind material seen projected against the stellar disk because of an azimuthal dependence of the mass-loss rate. Note in this respect that in HD 14134 maximum \has emission and maximum light do not seem to exactly coincide (see Fig.5). The existence of such a phase offset (which needs to be confirmed thanks to a more accurate ephemeris) would be consistent with a picture in which photospheric perturbations lead to the development of large-scale, spiral-shaped wind structures (Cranmer \& Owocki 1996). Let us suppose a bright photospheric region causing a local enhancement of the mass-loss rate. One observes a classical P Cygni profile when this `overloaded' portion of the wind is projected against the stellar disk. However, the motion of this wind feature towards the stellar limb because of rotation will cause the growth of \has emission. A small phase lag (i.e., $\Delta \phi$$<$0.25) would be expected if the spot has significant azimuthal extent. A substantial modulation of the local mass-loss rate may also be invoked to explain the large, phase-locked EWs variations seen in other stars (e.g., HD 13854; Fig.3).

\subsection{Relation to photospheric variability}
The hypothesis that the development of azimuthally structured outflows in early-type stars is intimately linked
to processes taking place at the photosphere (Cranmer \& Owocki 1996) has received considerable observational
support (e.g., Henrichs, Kaper \& Nichols 1994; Reid \& Howarth 1996). Much more controversial, however,
is the nature of the 'seed' perturbations, with pulsational and magnetic activity as the most straightforward
candidates. In the light of our results, we discuss below the relevance of these two scenarios in B-type
supergiants.

\subsubsection{Pulsations}
The {\it Hipparcos} data have revealed the existence of a large population of B-type supergiants with
daily, quasi-periodic light variations. These stars might constitute the extension to higher luminosities of the
slowly pulsating B stars, suggesting the $\kappa$-mechanism excitation of multiple, non-radial gravity modes
arising from a metal opacity bump at $T$ $\sim$ 2 $\times$ 10$^5$ K in stellar interiors (e.g., Waelkens \etal 1998). Both
the amplitude and the time-scale of the photometric variations are compatible to what is observed in our sample. To investigate this issue in more detail, we plot in Fig.7 our stars on a
HR diagram, along with the theoretical instability strips for non-radial g modes calculated {\it without} mass
loss and rotation (Pamyatnykh 1999). The stars with a clear cyclical behaviour (a sample evidently biased
towards the stars with large photometric amplitudes) do not appear to fall in the predicted domains of pulsational
instability. The NLTE spectral synthesis of Kudritzki \etal (1999) suggests systematically higher effective
temperatures and luminosities for some stars included in our study. However, such a temperature offset may not
be sufficient to bring these stars in the instability zone. Furthermore, only the late B-type stars in our sample
seem to follow the period-luminosity relation for g modes (see fig.3 of Waelkens \etal 1998). Massive stars are also believed to experience a variety of strange-mode and resonance instabilities
(Kiriakidis, Fricke \& Glatzel 1993), but there is no indication that the photometric
variations are related to this phenomenon (Fig.7). The ubiquity of pulsational instability in this part of the HR diagram (e.g., Fullerton \etal 1996; Kaufer \etal 1997) leads us to consider pulsations as most likely responsible for the photometric variations, although it is clear from the above discussion that the pulsation modes have yet to be identified in this case. The apparent dichotomy between the photometric properties of O and B supergiants seen in Fig.7 (the former exhibiting significantly lower variability amplitudes) suggests that the pulsations are of different nature in these 2 class of objects (e.g., higher modes are excited in O-type stars).  

\begin{figure*}
\epsfbox{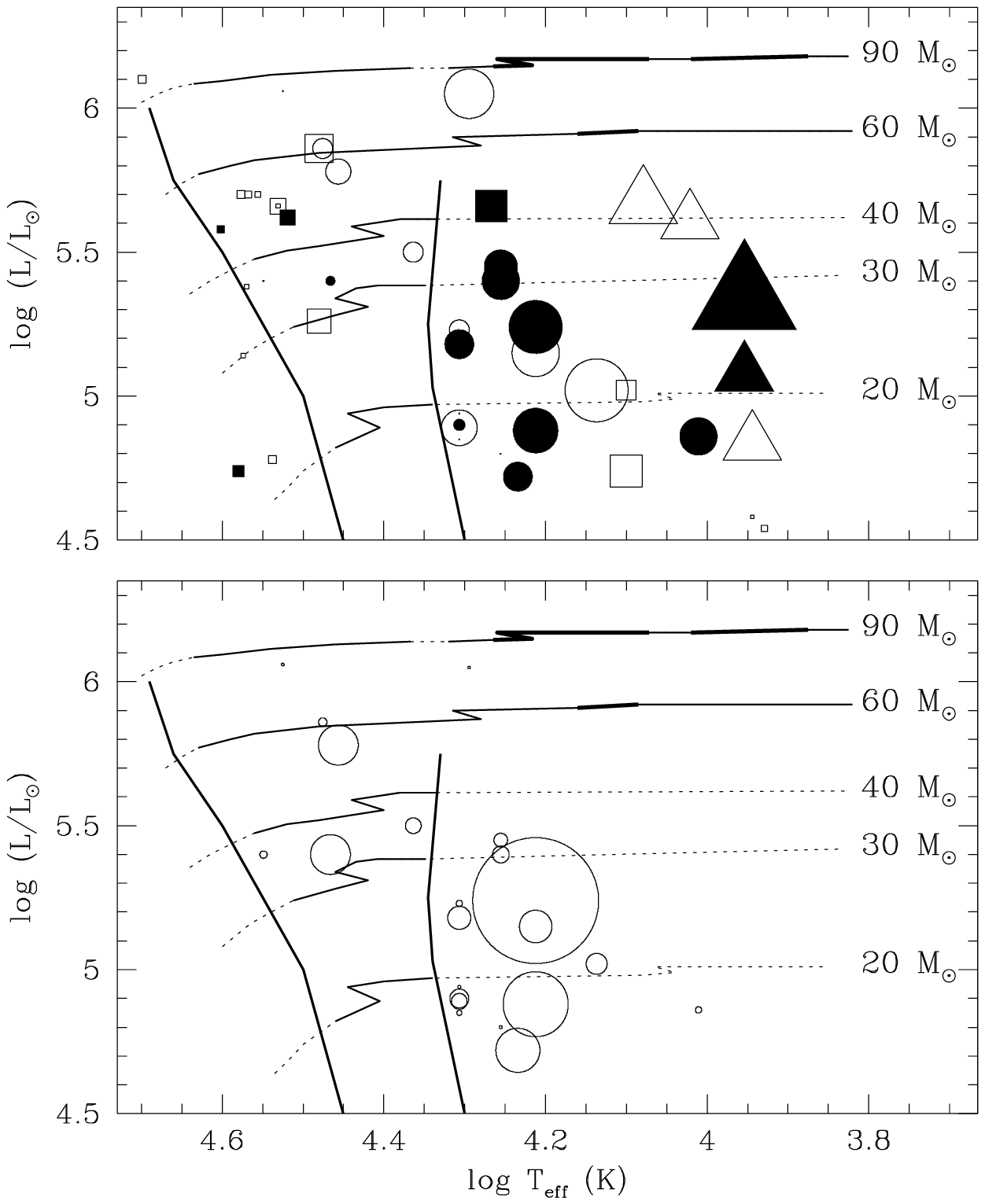}
\caption{{\it Upper panel}: the circles show the positions of the program stars in the HR diagram (some stars have been slightly shifted along the ordinate axis for the sake of clarity). The size of the symbols is directly proportional to the amplitude of the photometric variations in the {\it Hipparcos} data, $a_{\rm phot}$ (see Table 4). The squares and triangles show the position of galactic OBA-type stars (van Genderen \etal 1989) and blue supergiants in NGC 300 (Bresolin \etal 2004), respectively. The size of the symbols is proportional in this case to the amplitude of the variations in $V$-band. Filled and open symbols correspond to stars with and without evidence for photometric periodicities, respectively. {\it Lower panel}: positions of the program stars in the HR diagram with the size of the symbols being proportional to the fractional amplitude of the line-profile variations in \ha, $a_{\rm lpv}$. The evolutionary tracks ({\it dotted lines}) have
been calculated for solar metallicity and for the various initial masses indicated to the right-hand side
of the panels. Domains of the strange mode oscillations are overplotted as solid lines, while thick lines
correspond to very unstable phases (from Kiriakidis \etal 1993). The pulsational instability strip of non-radial
g-type modes is shown as (nearly vertical) solid lines. The models have been calculated using the OPAL opacities,
a hydrogen abundance, $X$=0.70, and a heavy element abundance, $Z$=0.02 (from Pamyatnykh 1999).}
\end{figure*}

The complete absence of the photometric periods in the \has time-series (HD 14134 excluded) seems at first glance to argue against pulsations as driver of the wind variability (once again under the assumption that the light variations can be ascribed to this mechanism). As can be seen in Fig.7, the stars displaying the highest level
of \has variability are not preferentially found close to the zones of pulsational instability. This hypothesis can be investigated further by looking for a correlation between the amplitude of the photometric
and \has changes, as parametrized by the activity indices $a_{\rm lpv}$ and $a_{\rm phot}$ (the latter
calculated from {\it Hipparcos} data; Section 4). Although there is a hint of a positive correlation
(FAP$\sim$2.3 per cent), this trend disappears when the 4 stars displaying line-profile variations
that cannot be unambiguously related to wind activity, but can be of purely photospheric origin, are excluded (i.e., HD 21291,
HD 24398, HD 31327, and HD 119608). In addition, there is also no correlation between $a_{\rm lpv}$ (\ha) and $a_{\rm lpv}$ (\he).

The lack of a clear statistical relationship between the levels of activity at
the photosphere and at the base of the outflow (as diagnosed by the photometric changes/\hes and \ha,
respectively) suggests that the same physical mechanism (i.e., presumably pulsations) is not responsible
for the photospheric and wind activity. This conclusion relies, however, on statistical grounds and should be
in principle examined on a star-to-star basis: here we note the example of HD 64760, where
the same periodic signal seems to appear in photospheric and wind lines (Kaufer \etal 2002). Although detailed models are still lacking, one may also expect a complex interplay between any pulsation signal and wind changes (e.g., beating of some pulsation modes). This is likely to result in a loose (if any) relationship between photometric and \has changes, for instance.  

\subsubsection{Magnetic fields}
A growing amount of evidence suggests the existence of weak magnetic fields in early-type stars, although firm
detections are still scarce and might be restricted to some atypical cases. Spectropolarimetric observations of
the very young O4--6 star $\theta$$^1$ Orionis C have revealed a relatively strong magnetic field, 1.1$\pm$0.1 kG
(Donati \etal 2002). The existence of a weaker field in the B1 IV star $\beta$
Cephei (with a polar component of 360$\pm$30 G) has also been reported recently. This field appears to be
sufficient to control the outflow at large distances from the star (out to $\sim$9$R_{\star}$; Donati \etal 2001).
Magnetic activity in OB stars is also hinted at by X-ray observations of the O9.5 II star HD 36486 ($\delta$ Ori)
and HD 37742 (Waldron \& Cassinelli 2000; Miller \etal 2002) which challenge the generally accepted view
that the X-ray emission in early-type stars solely arises from outwardly propagating shocks
developing as the
result of the inherently unstable nature of radiation-driven winds (see also Feigelson \etal 2002). Recent
theoretical studies indicate that a magnetic field of a moderate strength can be produced via dynamo action
at the interface between the convective core and the outer radiative envelope, and brought up to the
surface on a time-scale typically smaller than the main-sequence lifetime (Charbonneau \& MacGregor 2001;
MacGregor \& Cassinelli 2003).

Apart from our observations of HD 14134 (Section 5.2), HD 24912 is the only star in our sample
for which a magnetic field has been searched for via spectropolarimetric observations. A 3-$\sigma$
upper limit of 210 G in the longitudinal component was derived (de Jong \etal 2001). It is worth noting
that this does not set stringent constraints on the potential role played by magnetic activity, as even relatively weak fields ($B$$\la$100 G) are suspected to significantly perturb line-driven flows (Owocki 1994).
This issue has been recently worked out in more detail by ud-Doula \& Owocki (2002) who studied
the influence of a large-scale magnetic field with a dipole configuration
on the radiatively-driven outflow of an early-type star. They found that the disturbance imposed
by the magnetic forces can be parametrized by a confinement parameter,
$\eta$=$B^2_{\rm eq}$$R^2_{\star}$/$\dot{M}$$v_{\infty}$, where $B_{\rm eq}$ is the
equatorial field strength at the stellar surface. The magnetic field appears to significantly impact on
the wind morphology and dynamics even for values as low as $\eta$=0.1, while the emergent flow is channeled
along closed magnetic loops for $\eta$$\ga$1. For the physical parameters typical of our program stars (e.g., HD 14134;
see Table 1), we obtain $\eta$=1 for $B_{\rm eq}$$\sim$70 G. In this case the Alfv\'en radius would be about 2 $R_{\star}$, i.e., comparable to the extent of the \has line-formation region (ud-Doula \& Owocki 2002). This rough estimate shows that magnetic activity cannot be ruled out as driver of the wind activity at this stage. A much stronger
field is unlikely in HD 14134, considering our spectropolarimetric observations together with the lack of strong X-ray emission
($L_{\rm X}$[0.2--4.0 keV] $\la$ 33.0 ergs s$^{-1}$; Grillo \etal 1992). Fields of the same magnitude can also be expected in other stars from our sample in view of their relatively low X-ray luminosities (Bergh\"ofer, Schmitt \& Cassinelli 1996). 

\section{Conclusion}
Our long-term monitoring has confirmed the ubiquity of \has line-profile variations in OB supergiants. The changes result from the combined effect of a varying amount of incipient wind emission and an intrinsically variable underlying photospheric profile, with the relative contribution of these two phenomena varying from one star to another. As a consequence of the large amplitude of the \has variations observed, a precise determination of the stellar atmospheric parameters will not be generally possible from profile fitting of the Balmer lines in snapshot spectra. Taken at face value, for instance, the EW variations shown in Fig.3 would translate in the most extreme cases into a $\pm$0.8 dex uncertainty in the derived mass-loss rate (Puls \etal 1996; equation [43]). Although it remains to be seen whether a sophisticated modelling would result in drastically different figures, the large \has changes intrinsic to the early-type supergiants may be one of the most important factors limiting the accuracy with which these objects can be used as extragalactic distance indicators (see discussion in Kudritzki 1999). 

This survey has revealed the existence of cyclical \has line-profile variations in HD 14134 and HD 42087. 
The low detection rate (2 cases out of 17) may have several explanations, including observational limitations or a loss of coherency
in the patterns of variability. HD 14134 is of particular interest, as it is one of the very few early-type stars clearly exhibiting
the same pattern of variability in the photometric and spectroscopic modes. The existence of azimuthally structured outflows in our program stars might explain the dramatic \has changes,
but firm conclusions must await detailed modelling (see Harries 2000 for the first effort in this
direction). Sensitive searches for close companions (e.g., via long-baseline optical interferometric observations) are also needed to firmly rule out a variability arising from wind interaction.  

We do not find evidence for a direct coupling between the wind activity and base perturbations arising from pulsations, but we caution that our data are not well suited to examine this aspect in detail. It is conceivable that in several instances the globally anisotropic nature of the outflow primarily results from the putative existence of small-scale magnetic loops anchored at the stellar photosphere (or alternatively a presumably fossil, large-scale dipole field). A major observational effort is currently undertaken to detect such weak magnetic fields in early-type stars (e.g., Henrichs \etal 2003). Firm detections can be expected in the forseeable future. They would place the above conclusions on a quantitative footing and directly address the relevance of this scenario.

\section{Acknowledgments}
We are indebted to G. A. Wade and the TBL-MuSiCoS team for obtaining a single circular
polarization spectrum of HD 14134 during the mission 'Precision Studies
of Stellar Magnetism Across the HR Diagram' (July 2003). This work greatly benefited from the comments of the anonymous referee. The authors would like to acknowledge the support from the Applied Research and Technology 
Program at WKU and the assistance of the undergraduate telescope operators at WKU who participated in the 
observing program: A. Atkerson, T. Monroe and W. Ryle. It is a pleasure to thank the staff of the VBO for their excellent support. We are also grateful to the VBO chairman for the generous allotments of observing time. We wish to thank A. F. J. Moffat for useful comments. This research made use of NASA's Astrophysics Data System Bibliographic Services and the Simbad database operated at CDS, Strasbourg, France.

\bsp

\label{lastpage}

\end{document}